\documentclass[10pt,conference,letterpaper,onecolumn]{IEEEtran}

\usepackage{amsfonts,amsmath,amssymb}
\usepackage{amsthm}
\usepackage{graphicx}
\usepackage[caption=false,font=footnotesize]{subfig}
\usepackage{algorithm,algorithmicx,algpseudocode}
\usepackage{url}
\usepackage{comment}
\IEEEoverridecommandlockouts

\title{Content Caching and Delivery over Heterogeneous Wireless Networks}
\author{\IEEEauthorblockN{Jad Hachem}
\IEEEauthorblockA{University of California, Los Angeles\\
Email: jadhachem@ucla.edu}
\and
\IEEEauthorblockN{Nikhil Karamchandani}
\IEEEauthorblockA{Indian Institute of Technology, Bombay\\
Email: nikhilk@ee.iitb.ac.in}
\and
\IEEEauthorblockN{Suhas Diggavi}
\IEEEauthorblockA{University of California, Los Angeles\\
Email: suhas@ee.ucla.edu}
\thanks{This work was supported by the NSF grant \#1423271 and also through a gift from Qualcomm Inc.}
}

\newtheorem{theorem}{Theorem}
\newtheorem{lemma}{Lemma}

\newtheorem{corollary}{Corollary}
\newtheorem{example}{Example}
\newtheorem{definition}{Definition}

\newcommand{\floor}[1]{\left\lfloor{#1}\right\rfloor}
\newcommand{\ceil}[1]{\left\lceil{#1}\right\rceil}

\newcommand{\remove}[1]{}
\renewcommand{\epsilon}{\varepsilon}

\DeclareMathOperator*{\argmax}{argmax}

\usepackage{color}
\usepackage[usenames,dvipsnames]{xcolor}
\usepackage{soul}
\usepackage{cancel}

\newcommand{\mydelete}[1]{}

\newcommand{\myinsert}[1]{#1}

\newcommand{\myreplace}[2]{\mydelete{#1}\myinsert{#2}}

\newcommand{\mydeletem}[1]{}

\newcommand{\mycomment}[1]{}

\newcommand{\mycommentm}[1]{}

\newcommand{\myplotwidth}{.48\textwidth}

\newcommand{\myextendedref}[1]{Appendix~\ref{#1}}
\newcommand{\myextendedonly}[1]{#1}
\newcommand{\myshortonly}[1]{}

\begin{document}
\maketitle
\pagestyle{plain}

\begin{abstract}
Emerging heterogeneous wireless architectures consist of
a dense deployment of local-coverage wireless access points (APs)
with high data rates, along with sparsely-distributed,
large-coverage macro-cell base stations (BS).
We design a coded caching-and-delivery scheme for such architectures
that equips APs with storage, enabling content
pre-fetching prior to knowing user demands.  Users requesting content
are served by connecting to local APs with cached
content, as well as by listening to a BS broadcast transmission.  For any given
content popularity profile, the goal is to design the caching-and-delivery
scheme so as to optimally \myreplace{trade-off}{trade off} the transmission cost
at the BS \myreplace{with}{against} the storage cost at the APs and the user cost
of connecting to multiple APs. We design a coded caching
scheme for non-uniform content popularity that dynamically
allocates user access to APs based on requested content. We
demonstrate the approximate optimality of our scheme with respect to
information-theoretic bounds. We numerically evaluate it on
a YouTube dataset and quantify the
trade-off between transmission rate\mydelete{s}, storage, and access cost. Our
numerical results also suggest the intriguing possibility that, to
gain most of the benefits of coded caching, it suffices to divide
the content into a small number of popularity classes.

\end{abstract}

\section{Introduction}\label{sec:intro}
Broadband data consumption has witnessed a tremendous growth over the past few years, due in large part to multimedia applications such as Video-on-Demand.
This increased demand has been managed in the wired internet via Content Distribution Networks (CDNs), by mirroring data in various locations and in effect pushing the content closer to the end users.
Wireless data consumption, driven by the increased demand for high-definition content on mobile devices, has also grown at a significant rate \cite{CiscoReport} and is testing the limits of our underlying wireless communication systems  \cite{QualcommSmallCells}.
However, simply borrowing the CDN solution from wired networks and applying it to wireless systems is insufficient to solve the wireless content delivery problem.
In this work, we propose a content caching-and-delivery scheme, based on a new \emph{multi-level} storage, access, and distribution design, that is matched to the architecture of emerging wireless systems.
We analytically demonstrate the efficiency of our proposed scheme and numerically evaluate its performance on YouTube data. 

There are various reasons for why the traditional CDN architecture is not sufficient to solve the wireless content distribution problem.
In wired networks, CDNs carefully evaluate what content to store based on user demand, and then replicate popular content at several locations.
This helps reduce the load at the host server, by serving many user requests locally via the content cached in the local storage.
This solution works best when neither the local storage nor the data rates are a bottleneck\myextendedonly{ \cite{korupolu1999}}.
In typical wireless cellular systems, neither of these conditions are true: the wireless last-hop link is a bottleneck and the storage available at the cellular base-stations is limited. 

While there have been tremendous improvements in the cellular data rates over successive generations of wireless systems, the gains are not sufficient to compensate for the exploding rise in data demand.
This has led to the emergence of a heterogeneous wireless network (HetNet) architecture, as a front-runner for 5G systems \myshortonly{\cite{QualcommSmallCells}}\myextendedonly{\cite{IntelHetNet, IntelSmallCells, QualcommSmallCells}}.
This architecture advocates a dense deployment of wireless access points (APs), with low coverage radius and high data rates, combined with a sparse deployment of macro-cellular base stations (BSs), with larger coverage but limited communication rates.
For example, the APs could be WiFi access points or small-cells, which help offload some of the macro-cellular data traffic.
However, the APs themselves do not provide a reliable solution since they are connected to the rest of the network via best-effort backhaul,
which is a bottleneck\myextendedonly{ \cite{IntelSmallCells}}.
It has been noted that even a joint management of APs and BSs is not sufficient to deal with the projected growth
in users and demand for data over wireless networks\myshortonly{ \cite{QualcommSmallCells,CiscoReport}}\myextendedonly{ \cite{IntelSmallCells, QualcommSmallCells, CiscoReport}}.

Given the above discussion, we note that both the traditional CDN approach as well as enhanced wireless system design are solving only one aspect of the wireless content delivery problem.
CDNs optimize content placement, without due consideration to characteristics of wireless communications.
On the other hand, wireless system design only focuses on increasing delivery rates, agnostic to content.
In this paper, we propose a joint design of content placement, access, and delivery for the heterogeneous wireless network architecture to address the wireless content delivery problem.
The key components of our design are:  {\sf (1)} \emph{Multi-level popularity}: Divide the content into different levels of popularity based on statistical knowledge of user requests; {\sf (2)} \emph{Multi-level caching}: Equip APs with storage capabilities and use them to locally cache content based on popularity; {\sf (3)} \emph{Multi-level access}: Dynamically allocate user access to APs, based on popularity of requested content; and {\sf(4)} \emph{Broadcast delivery}: Use the inherent physical-layer broadcast 
property of wireless communications to serve multiple (distinct) requests simultaneously, by enabling coded-multicasting opportunities.%
\footnote{The idea of coded caching was first proposed in \cite{maddah-ali2012, maddah-ali2013} for the case of uniform file popularity; see Section~\ref{sec:related} for a comprehensive overview of the related literature. In non-uniform settings, coded caching has been shown to perform better than traditional methods such as LFU \cite{niesen2013}.}
For any given content popularity profile, this design trades off the transmission cost at the BS with the storage cost at the APs, and the access cost at the users resulting from connecting to multiple APs.%
\footnote{The AP deployment envisioned in next-generation wireless networks (see for example \cite{QualcommSmallCells}) is dense enough such that mobile devices would often have multiple APs in their range.}

The main theoretical contribution of this work aims at solving this trade-off.
It consists of the design and analysis of a multi-level caching-and-delivery scheme for any given content popularity profile, available AP storage, and user access to the APs.
%In order to solve this trade-off, the main theoretical contribution of this work is the design and analysis of a multi-level caching-and-delivery scheme for any given content popularity profile, available storage at the APs, and user AP access structure.
The basic idea of the scheme is to divide the available storage at each AP among the various popularity levels (\emph{memory-sharing}), and then cache content so as to create the maximum number of coded-multicasting opportunities during delivery.
By comparing the achievable BS transmission rate of our scheme to information-theoretic lower bounds,%
\footnote{The information-theoretic lower bounds do not assume
memory-sharing placement, or any other placement or delivery scheme.}
we are able to demonstrate its order-optimality for any content popularity profile, available storage, and user access structure.
One of the main technical innovations of the paper is the introduction of new non-cut-set-based information-theoretic lower bounds, which are critical to the proof of the order-optimality of our proposed scheme for the multi-level, multi-access system.
\myshortonly{The complete proof is quite long, and we are unable to include it here due to shortage of space.
We will discuss the main ideas of the proof here and refer the reader to \cite{extended} for all the details.}

A striking aspect of the proposed near-optimal scheme is that, in some regimes, we choose to store some content of lower popularity even though more popular content has not been completely cached.
Another unique aspect is that we allow users to have different access structures, based on the popularity level of the content that they request.
Our results demonstrate the significant benefits in performance that can be derived by enabling such multi-level access. 

We also do an evaluation of our proposed scheme on a YouTube dataset that is based on user requests \cite{YoutubeRepository}.
This dataset provides the
``continuous'' content popularity profile seen in \figurename~\ref{fig:youtube}.
We begin by clustering the content into different levels,
using efficient heuristic methods whose performance is very close to that of expensive brute-force techniques,
and then evaluate the performance of our proposed multi-level caching-and-delivery scheme. 
The numerical results suggest the intriguing possibility that, to gain most of the benefits of coded caching-and-delivery, it suffices to divide the content into at most three or four levels of popularity.
Finally, these numerical evaluations also provide insight into how the user access structure impacts performance.
For example, we find that, in certain regimes, it is better to require users requesting more popular content to access a single AP and those requesting less popular content to connect to more APs. 

The paper is organized as follows.
In Section \ref{sec:model},
we formally state the problem and establish the notation used
throughout the paper.
Section \ref{sec:results} gives an overview of our main results and also discusses their implications.
We give some preliminary results from the literature in Section~\ref{sec:preliminaries},
which we will use in Section~\ref{sec:achievability} to
describe the multi-level
caching-and-delivery scheme.
We establish the approximate optimality of the scheme in Section~\ref{sec:converse}.
Section~\ref{sec:discussion} discusses extensions of our results beyond the analytical model described before to more realistic scenarios,
and numerically evaluates the scheme in these scenarios on real data traces.
Finally, we place our work in
context of the literature in Section \ref{sec:related}.
\myshortonly{Due to lack of space, we skip the detailed proofs here and instead provide them in \cite{extended}.}
\myextendedonly{All of the detailed proofs are relegated to the appendices.}

\remove{
In general, some files on a host server are much more likely to be requested than other files.
As an example, some YouTube videos have a few hundred views, while other, much more popular videos, have over a billion.
Such popularities are often modeled by a power law.
For instance, \figurename~\ref{fig:youtube} shows the popularity (in terms of number of views) of about half a million YouTube videos, as well as a Zipf distribution that approximates these popularities well.
Such popularity distributions can be used to model user demands for various files.

\begin{figure}
\centering
\includegraphics[width=.48\textwidth]{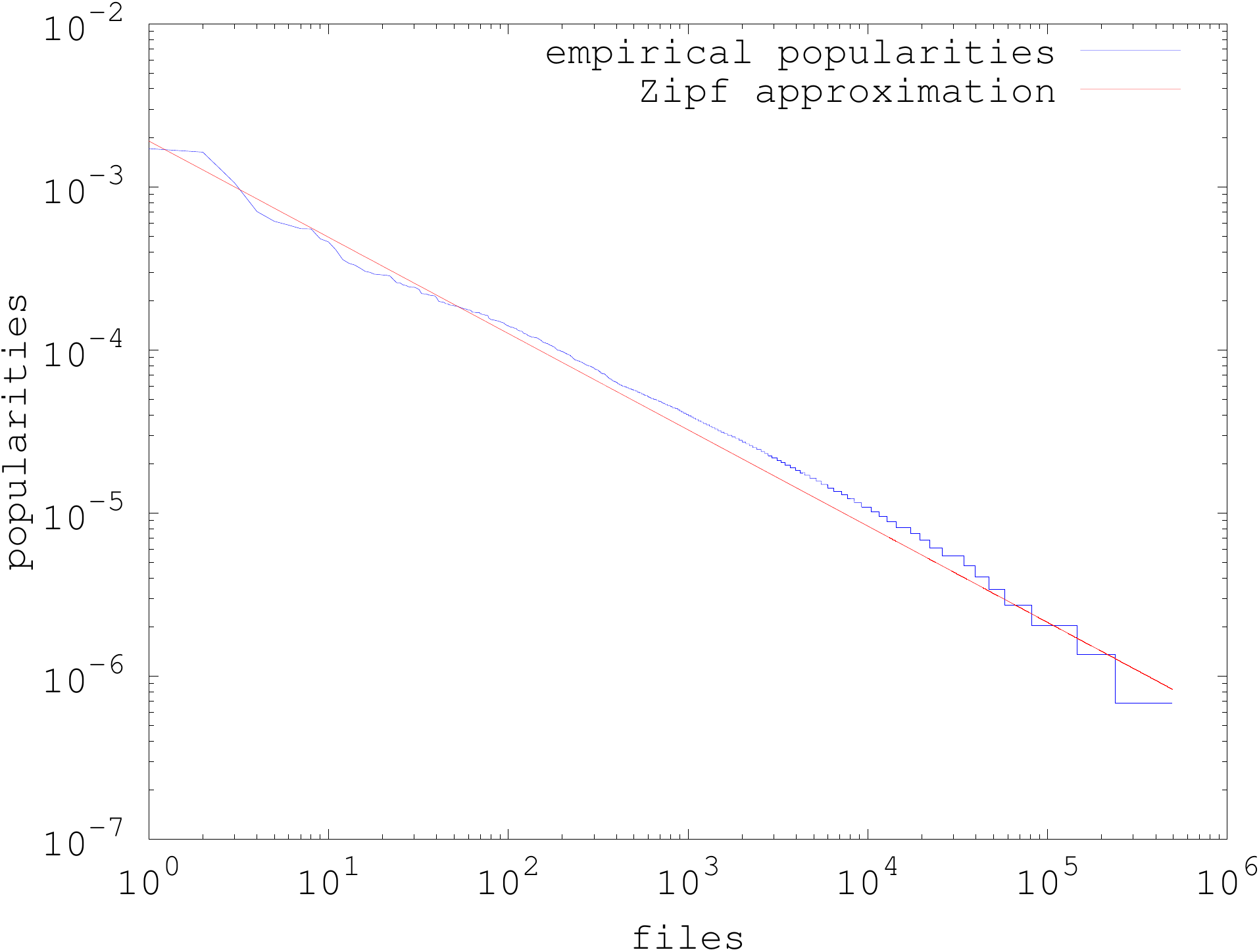}
\caption{Popularties, expressed in normalized number of views, of $493\,722$ YouTube videos.
The red line is a Zipf distribution with parameter $\approx0.6$ that best approximates these popularities, in an $L^2$ sense.}
\label{fig:youtube}
\end{figure}

As users enter the network, they connect to one or more access points and independently request files from the server, according to the popularity distribution of the files.
The server then sends a broadcast message, through the base station, to all the users in the network.
After receiving the broadcast message, each user combines it with the contents of their nearby caches to recover the file that they have requested.
The goal is to minimize the expected size of the costly broadcast message, for any access point cache memory.

\emph{[Mention different file sizes]}

As can be seen when searching for a Wi-Fi network on any device, a single device often has several access points in its vicinity.
If these access points are equipped with caches, then it is natural to consider having some of the users connect to more than one access point.
While this might increase the delay at these individual users, it could also create a greater benefit to the whole network by shifting the burden of content delivery from the base station to the access points.

In this paper, we first analyze a deterministic model that assumes files are divided into a small number of \emph{popularity levels}, which we present in Section~\ref{sec:model}.
In Section~\ref{sec:results}, we establish the main analytical results of the paper, which consist in an achievable rate that is approximately optimal for the deterministic model.
We describe the scheme that achieves this rate in Section~\ref{sec:achievability} and prove its approximate optimality in Section~\ref{sec:converse}.
Section~\ref{sec:discussion} discusses extending the scheme for the deterministic model to the general, stochastic model, as well as some other unresolved questions.
Finally, we numerically evaluate the scheme on the stochastic model in Section~\ref{sec:numerics} and discuss the results.
}
%In this paper, we first simplify this general model to a more deterministic one, which we present in Section~\ref{sec:model}.
%In Section~\ref{sec:results}, we propose a scheme that is order-optimal for this simplified model.
%We then numerically evaluate its performance on the general model in Section~\ref{sec:numerics}, and discuss the results and future improvements in Section~\ref{sec:discussion}.

\section{Problem setup and notation}\label{sec:model}

%\subsection{The deterministic setup}
\label{sec:model-deterministic}

\begin{figure}
\centering
\includegraphics[width=.38\textwidth]{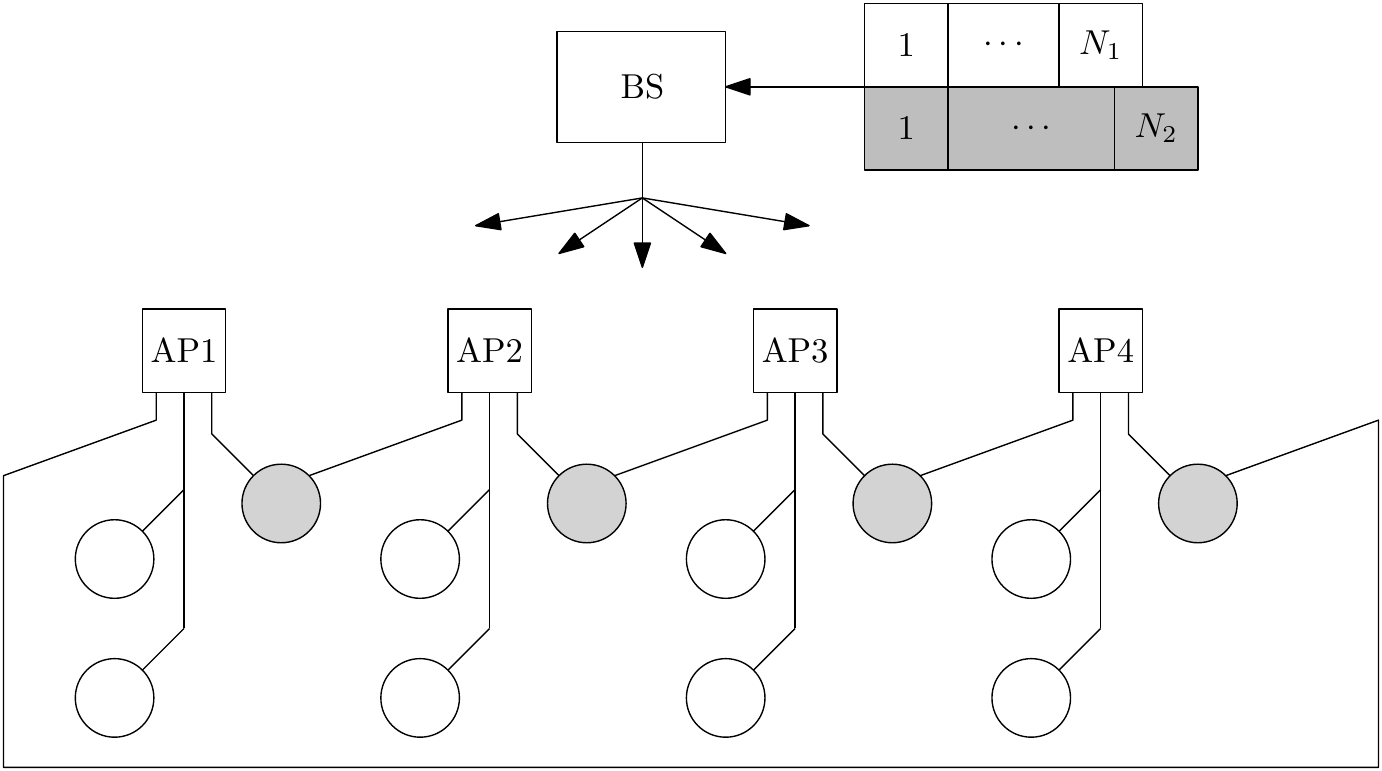}
\caption{Setup with $L=2$ levels.
There are $K=4$ caches/access points.
Level 1 has $N_1$ files and $U_1=2$ users per cache (in white), and these users are required to access only $d_1=1$ cache.
Level 2 has $N_2$ files and $U_2=1$ user per cache (in gray), and these users are required to access $d_2=2$ consecutive caches.
Notice that the last user accesses caches $4$ and $1$, for symmetry.}
\label{fig:setup}
\end{figure}

Consider the caching system illustrated in \figurename~\ref{fig:setup}.
The server hosts $N$ files of size $F$ bits each.
Each file belongs to one of $L$ popularity levels, labeled $1$ through $L$, such that the total number of files in level $i\in\{1,\ldots,L\}$ is $N_i$ files.
The network consists of $K$ access points, equipped with caches that can store up to $M\ge0$ files, or, equivalently, $MF$ bits.
There are $U$ users per cache in the network, for a total of $KU$ users in the system.
Of the $U$ users per cache, exactly $U_i$ users request files from popularity level $i$, and these users are required to access content from the caches of the $d_i$ closest access points.
We refer to $d_i$ as the \emph{access degree} of the level-$i$ users.
We will assume that there is a small constant $D$ such that $d_i \le D$ for all popularity levels $i$. 
%Naturally, it is impractical to require that some users connect to a very large number of access points, so we enforce the condition that $d_i\le D$ for some (small) $D$.
For symmetry, we will also assume that the access points are arranged in a cyclic fashion, such that cache $1$ and cache $K$ are next to each other.
%Thus, the user profile is exactly the same when viewed from every access point.

%In our setup, the popularity of a file $W$ from level $i$ is proportional to $U_i/N_i$, since this quantity represents the average fraction of users requesting $W$.
In our setup, the popularity of a file $W$ from level $i$ is proportional to the average fraction of users requesting $W$, which is $U_i/N_i$.
Without loss of generality, we assume:
\begin{equation}\label{eq:level-ordering}
\frac{U_1}{N_1} \ge \frac{U_2}{N_2} \ge \cdots \ge \frac{U_L}{N_L}.
\end{equation}
Thus, files from level $1$ are the most popular, while level $L$ contains the least popular files.

\remove{
The popularity of a level can be directly determined from the parameters.
By definition, the popularity of a file is the fraction of users that are requesting it.
In this setup, the popularity of a file from level $i$ is thus proportional to $\frac{KU_i}{N_i}$, since there is a total of $KU_i$ users for all $N_i$ files in level $i$.
Therefore, the most popular level is the one with the largest $\frac{U_i}{N_i}$, and the least popular level is the one with the smallest $\frac{U_i}{N_i}$.
Thus, level $1$ is the most popular level, and level $L$ is the least popular one.
}

The system is required to operate in two phases.
In the first phase, called the \emph{placement phase}, information about the files is stored in the caches of the $K$ access points.
%The contents of cache $k$ are denoted by $Z_k$.
This phase occurs prior to any knowledge about user requests.
The second phase, called the \emph{delivery phase}, takes place after the $KU$ users reveal the files that they have requested.
The base station transmits a single broadcast message of size at most $RF$ bits, which all users can hear.
The users must then use the broadcast message, together with the contents of the caches that they can access, to recover the files that they have requested.

Henceforth, we will refer to $R$ as the BS transmission rate and $M$ as the memory size of each cache.
A rate-memory pair $(R,M)$ is said to be \emph{achievable} if there exists a placement strategy in caches of capacity $MF$ bits such that, for any possible user request profile, a broadcast message of size $RF$ bits can be delivered to satisfy all the requests.
For any memory size $M \ge 0$, the optimal BS transmission rate is:
\begin{equation}
\label{Eq:optimal-rate}
R^\ast(M) = \inf\left\{ R\ge0 : \text{$(R,M)$ is achievable} \right\},
\end{equation}
where the minimization is over \emph{all} possible caching and delivery schemes.
Given the problem parameters $\{N_i, U_i, d_i\}_i$ and $K$, our goal is to characterize this optimal rate $R^\ast(M)$ for each value of memory size $M$. 
\remove{
Both the size of the memory and that of the broadcast message are costly.
Therefore, a natural problem is to find the trade-off between these two quantities.
We denote by $M$ the size of each cache, normalized by the file size $F$ (\emph{i.e.}, the cache can hold $MF$ bits).
Similarly, we denote by $R$ the size of the broadcast message, normalized by the file size $F$.
These normalized quantities are referred to as \emph{memory} ($M$) and \emph{rate} ($R$).

We wish to characterize the memory-rate trade-off required to serve all requests, even in the worst case.
In particular, a rate-memory pair $(R,M)$ is said to be \emph{achievable} if there exists a placement strategy in caches of capacity $MF$ bits such that, for any possible user request vector $(r_1,\ldots,r_{KU})$, a broadcast message of size $RF$ bits can be delivered to satisfy all the requests, with vanishing error probability as $F$ grows.
We formulate the problem as finding the smallest achievable $R$ for every $M$ as in the next definition.
\begin{definition}\label{def:optimal-rate}
For any cache memory $M\ge0$, the optimal broadcast rate is defined as:
\[
R^\ast(M) = \inf\left\{ R\ge0 : \text{$(R,M)$ is achievable} \right\}.
\]
\end{definition}

We stress that the minimization in Definition~\ref{def:optimal-rate} is done across \emph{all possible strategies}.
In other words, the rate $R^\ast(M)$ is the smallest possible achievable rate, irrespective of any particular choice of caching-and-delivery strategy.
}

\myextendedonly
{
For convenience, we summarize the notation used in this paper in Table~\ref{tbl:notation}.
\begin{table}
\renewcommand{\arraystretch}{1.3}
\centering
\caption{Notation}
\label{tbl:notation}
\begin{tabular}{|c|l|} 
  \hline
    $N$ & Total \#\ of files \\
  \hline
    $L$  & \#\ of file classes \\
  \hline
    $N_i$ & \#\ of files in the $i^\text{th}$ popularity level \\
  \hline
    $K$ & \#\ of APs \\
  \hline
    $M$ & Memory size of cache at each AP \\
  \hline
    $U$ & \#\ of users requesting files from an AP \\
  \hline 
    $U_i$ & \#\ of users per AP requesting $i^\text{th}$ level files\\
  \hline
    $d_i$ & \#\ of AP caches accessed by a user requesting an $i^\text{th}$ level file\\
  \hline
    $D$ & Maximum $d_i$ over all $i$\\
  \hline
    $R$ & Rate of BS transmission \\
  \hline
\end{tabular}
\end{table}
}
\remove{
The setup studied in this paper is an extension of the one in \cite{MLcodedcaching}, which dealt with users accessing only a single cache each.
}

\subsection{Regularity conditions}\label{sec:model-regularity}

We now present some regularity conditions that naturally arise in practice and that simplify the theoretical analysis.
First, we assume that, for each popularity level, there are more files than users requesting those files at any given time:%
\footnote{For example, this is easily seen to be true in video applications such as Netflix, where each ``file" corresponds to a short segment of  video of length ranging from a few seconds to a minute.
If there are a $1000$ popular files, each of length $60$ minutes, this would correspond to more than $60\,000$ files.}
%In other words,
\begin{equation}\label{eq:regularity-more-files-than-users}
\forall i\in\{1,\ldots,L\},\quad N_i\ge KU_i.
\end{equation}

Second, we will assume that no two levels are very close in popularity.
Indeed, if two levels had similar popularities, then they can be combined into a single level.
%As we mentioned before, the popularity of a file is proportional to $U_i / N_i$ in our setup.
%Therefore, the regularity condition can be expressed as:
We express this by:
\begin{equation}\label{eq:regularity-level-separation}
\forall \ i<j,\quad \frac{U_i/N_i}{U_j/N_j} \ge q,%= \frac{256(D+1)^4L^2}{(1-e^{-1})^2}.
\end{equation}
for some constant $q>1$, since $U_i/N_i$ represents the popularity of level $i$.
Recall that, from \eqref{eq:level-ordering}, $i<j$ implies that level $i$ is more popular than level $j$.
\remove{
Finally, for simplicity, we will assume that $K$ is a multiple of $d_i$ for all levels $i$:
\begin{equation}\label{eq:regularity-k-multiple-di}
\forall i\in\{1,\ldots,L\},\quad d_i\mid K.
\end{equation}
This simplifies the scheme that we use as well as the analysis.
If $d_i\nmid K$, then a small number of the caches ($K\bmod d_i$) would suffer from some edge effect, which does not significantly affect the overall performance.
}

\section{Main theoretical results}\label{sec:results}

The theoretical results presented in this paper are two-fold.
We first propose a caching-and-delivery scheme for our setup with multi-level popularity and access structures.
For future reference, we call this scheme Multi-Level Popularity-Aware Memory Allocation (ML-PAMA).
The scheme is a non-trivial generalization of the one proposed for single-access, multi-level caching in \cite{MLcodedcaching} to multi-access systems.
The main technical difficulty here comes from the correlations introduced in the cached content available to different users in the system, due to partial overlaps between the accessed APs.
Our proposed scheme manages to remove these correlations by carefully dividing the users and caches into different groups so that there are no overlaps within a group, and then serving the different groups separately.
We validate this design by proving that our proposed scheme is in fact \emph{order-optimal}, \emph{i.e.}, for any given problem parameters $K$ and $\{N_i, U_i, d_i\}$, and any memory size $M$, the BS transmission rate $R(M)$ achieved by our scheme is within a constant gap of the information-theoretically optimal rate $R^\ast(M)$ defined in \eqref{Eq:optimal-rate}.
The key technical contribution here is the introduction of new non-cut-set information-theoretic lower bounds for the multi-level, multi-access system, which are partly based on the \emph{sliding-window subset entropy inequality} \cite{Liu14}, and their use in the proof of the order-optimality of our proposed scheme ML-PAMA.

\subsection{Caching-and-delivery scheme (ML-PAMA)}\label{sec:results-achievability}

In \cite{maddah-ali2013}, an order-optimal scheme was given for a special case of this problem in which there is only a single popularity level.
This scheme randomly and independently places content in the caches, and then sends coded broadcast messages that can benefit subsets of users at once, by taking advantage of coding opportunities.
%The coding (linear combinations) in each BS transmission is beneficial particularly because the cache contents accessed by the different target users are independent.
%When users have access to the same cache side-information, fewer coding opportunities are available.
In our multi-level setup, we propose to separate the levels, both in the AP caches and in the BS transmission, thus resulting in $L$ single-level subsystems.
Since every level is represented by users at every cache, we have enough coding opportunities within each level, suggesting that coding across levels is not required.
%In our multi-level setup, for every popularity level there are users accessing each cache.
%Hence, there are always coding opportunities between users of the same level, and thus there is no need to look for coding across levels.
%With that in mind, we propose to separate the levels in the multi-level system, both in the AP caches and in the BS transmission, thus resulting in $L$ single-level subsystems.
The separation allows us to allocate different parts (of varying sizes) of the memory to different levels.
Each part $i$, of size $\alpha_iM$ with $\alpha_i\in[0,1]$ and $\sum_i\alpha_i=1$, is allocated to level~$i$.
This allocated portion $\alpha_iM$ is exclusively used for storing content related to files from level $i$.
When, later in the delivery phase, the users request various files from the server, the base station sends $L$ separate broadcasts, one to serve the requests from each popularity level.
Then, the total BS transmission rate is given by:
\begin{equation}\label{eq:total-rate}
R\left(M,K,\{N_i,U_i,d_i\}_i\right) = \sum_{i=1}^L R^\text{SL}\left(\alpha_i M, K, N_i, U_i, d_i\right),
\end{equation}
where $R^\text{SL}$ denotes the achievable rate of a single-level system.
Thus $R_i(M) = R^\text{SL}(\alpha_iM,K,N_i,U_i,d_i)$ is the individual rate of the message serving level $i$.
For ease of notation, we will henceforth denote the total transmission rate as $R(M)$.

For each of the $L$ levels, we generalize the coded caching-and-delivery scheme proposed in \cite{maddah-ali2013} to the case where users can access multiple caches.
As mentioned before, the main challenge here was dealing with the correlations introduced by the partial overlap between the caches accessed by different users; we use a coloring-based approach to take these into account. 
Next, we identify the appropriate choice for $\{\alpha_i\}$, which we call the \emph{memory-sharing parameters}, so that the total achievable rate $R(M)$ in \eqref{eq:total-rate} is minimized.
This gives us our first result, summarized in the following theorem.
\begin{theorem}\label{thm:multi-level-achievability}
For the multi-level caching setup with $L$ levels, $K$ caches, and, for each level $i$, $U_i$ users per cache with access degree $d_i$ and $N_i\ge KU_i$ files, the following rate is achievable for any $M\ge0$:
\[
R(M) = \sum_{h\in H}KU_h + \frac{ \left( \sum_{i\in I}\sqrt{N_iU_i} \right)^2 }{ M - \sum_{j\in J}N_j/d_j } - \sum_{i\in I} d_iU_i,
\]
where $(H,I,J)$ is an $M$-feasible partition of the set of levels.%
\footnote{An $M$-feasible partition of the set of levels is a partition for which the cache memory $M$ satisfies a certain set of inequalities.
See Definition~\ref{def:m-feasible} in Section~\ref{sec:achievability-memory-sharing} for more details.}
\end{theorem}

Details of the scheme are provided in Section~\ref{sec:achievability}, and the full proof of the theorem is given in \myextendedref{app:achievability}.

\subsection{Order-optimality of the scheme}\label{sec:results-converse}

The memory-sharing scheme that we discussed above is one of many possible strategies to solve the caching-and-delivery problem in our setup.
Our next result shows that it performs approximately as well as the information-theoretic optimum, for all values of the problem parameters.
\begin{theorem}\label{thm:multi-level-optimality}
The rate $R(M)$ achieved by ML-PAMA is within a multiplicative factor of the information-theoretically optimal rate $R^\ast(M)$.
This factor is upper-bounded by $37(D+1)^3L^3$.
\end{theorem}

Notice that the gap depends on both the maximum access degree $D$ and the number of levels $L$, but is independent of all other problem parameters.
One would expect $D$ to not be very large, due to the delay associated with connecting to multiple APs and retrieving data from their caches.
As for $L$, our numerical evaluations in Section~\ref{sec:numerics} suggest that, in practice, dividing the files into a small number of levels is enough to derive most of the caching gains.
Therefore, $L$ will likely also be a small constant, thus making the whole multiplicative factor also a constant.

Furthermore, the multiplicative gap in Theorem~\ref{thm:multi-level-optimality} is very generous, and was chosen so for the sake of simplifying the analysis.
In fact, we expect the dependence of the gap on $D$ and $L$ to be much weaker than the $O(D^3L^3)$ dependence indicated above; we believe it can be reduced to $O(D)$.%
\footnote{We were recently able to completely remove the dependence on $L$ in the case $D=1$. We believe this improvement extends to $D>1$, and all these results would be reported in \myshortonly{\cite{extended}}\myextendedonly{this document}.}
As we will see in Section~\ref{sec:numerics-gap}, numerical results strongly support our intuition that the true gap is in fact much smaller than what the theoretical results indicate.

\section{Preliminaries}\label{sec:preliminaries}
Our solution for the multi-level popularity and user access structures is based on the coded caching-and-delivery scheme developed for the single-level, single-user, single-access setup in \cite{maddah-ali2013}.
We illustrate here the main ideas of the scheme, as well as its associated rate, via an example.

\begin{example}[Coded caching example \cite{maddah-ali2013}]\label{ex:coded-caching}
Consider $K=2$ caches, $L = 1$ level, $N_1 = N$ files, and $U_1 = 1$ user per cache with access to $d_1 = 1$ cache.
The placement phase consists of storing $MF/N$ independently and randomly sampled bits of every file in each cache.
Each file $W^n$ can then be seen as being composed of four parts:
\(
W^n = \left(W^n_0, W^n_1, W^n_2, W^n_{1,2} \right)
\),
such that $W^n_{\mathcal{S}}$ denotes the bits of file $W^n$ which are exclusively stored in the caches in the set $\mathcal{S}\subseteq\{1,2\}$.%
\footnote{For ease of notation, we write $W^n_{1,2}$ instead of $W^n_{\{1,2\}}$, etc.}
For example, $W^n_1$ represents those bits that are stored in cache~1 but not in cache~2.
Since the bits are independently and randomly sampled, the size of each file part is:
\begin{equation}
\label{Eqn:SizeParts}
|W^n_{\mathcal{S}}| \approx \left(M/N\right)^{\left|\mathcal{S}\right|} \left( 1 - M/N\right)^{2 - \left|\mathcal{S}\right|}F.
\end{equation}

In the delivery phase, suppose users 1 and 2 (connected to caches 1 and 2, resp.) request files $W^1$ and $W^2$, resp.
Then, the server will send the following broadcast:
%\begin{equation}\label{eq:example-broadcast}
\(
\left(W^1_0\,,\,W^2_0\,,\,W^1_2 \oplus W^2_1\right)
\),
%\end{equation}
where ``$\oplus$'' denotes the bit-wise XOR operation.
Using \eqref{Eqn:SizeParts}, the rate of this transmission is:
\begin{IEEEeqnarray*}{rCl}
R &=& 2 \left( 1 - M/N \right)^2 + (M/N) \left(1 - M/N \right)\\
&=& \left(N/M - 1 \right) \cdot \left( 1 - \left(1 - M/N \right)^2 \right). \IEEEyesnumber \label{eq:example-broadcast-rate}
\end{IEEEeqnarray*}

User 1 now has $W^1_1$ and $W^1_{1,2}$ from the contents of cache~1, and also $W^1_0$ from the broadcast.
Furthermore, they can also recover $W^1_2$ by combining $\left(W^1_2\oplus W^2_1\right)$ transmitted by the server with $W^2_1$ stored in cache~1.
%Furthermore, from $\left(W^1_2\oplus W^2_1\right)$ transmitted by the server and using $W^2_1$ stored in cache~1, user~1 can also recover $W^1_2$.
Thus, by accessing the contents of cache~1 and listening to the BS transmission, user~1 can recover the file $W^1 = (W^1_0, W^1_1, W^1_2, W^1_{1,2})$.
Similarly, user~2 can recover file $W^2$.
\end{example}

The scheme can be generalized to an arbitrary $K$, and the associated rate is given by the following lemma.
\begin{lemma}[Single-level, single-user, and single-access achievable rate \cite{maddah-ali2013}]\label{lemma:single-level-user-access}
For $L=1$ level, $K$ caches, a single user connecting to each cache (and to no other), $N\ge K$ files, and memory size $M$, the following BS transmission rate is achievable: 
\[
\left[  \left(N/M-1\right)\cdot\left(1 - \left(1-M/N\right)^K\right) \right]^+,
\]
where $[x]^+ = \max\{x,0\}$.
%Furthermore, the rate $r(0,K,N)=K$ can be achieved when $M=0$.
\end{lemma}

For example, when $M = N$ and $M = 0$, the BS transmission rates are $0$ and $K$ respectively. 
\remove{
The placement phase consists in sampling a random subset of size $\frac{MF}{N}$ of each file in each cache, exactly as in the $K=2$ example.
Likewise, the delivery phase consists in transmitting a network-coded broadcast that the users can decode using the contents of their cache.

By analyzing the expected size of each part of $W^n$, it is possible to calculate the expected size of the broadcast.
For large file sizes, the actual broadcast size will be very close to the expected size with high probability.
The following lemma gives this expected size.
}
% remove
\remove{
The expression of $r(\cdot)$ can be simplified by upper-bounding it as:
\[
r(M,K,N) \le \left(1-\frac{M}{N}\right) \cdot \min\left\{ \frac{N}{M} , K \right\},
\]
for all $M\in[0,N]$.
}

\section{Caching-and-delivery strategy}\label{sec:achievability}
As discussed in Section~\ref{sec:results-achievability}, our proposed caching-and-delivery scheme divides the multi-level system with $L$ popularity levels into $L$ independent single-level subsystems.
This section elaborates first on the design of the scheme for a single-level subsystem with a given user access structure, and then on the optimal memory sharing between the $L$ levels. 
\remove{
The caching scheme described in Section~\ref{sec:results-achievability} divides the original problem into two subproblems.
The first is the design of a scheme for a single-level system, and the second is the design of the memory-sharing parameters $\alpha_i$.
This section elaborates on how to solve these subproblems.
}
\subsection{Single-level with multi-user and multi-access}\label{sec:achievability-single-level}

Consider a single-level system consisting of $N$ equally popular files, $K$ caches, and $U$ users per cache, each connecting to the $d$ closest caches.
Below, we describe an order-optimal scheme for such a system.

\begin{figure}
\centering
\includegraphics[width=.38\textwidth]{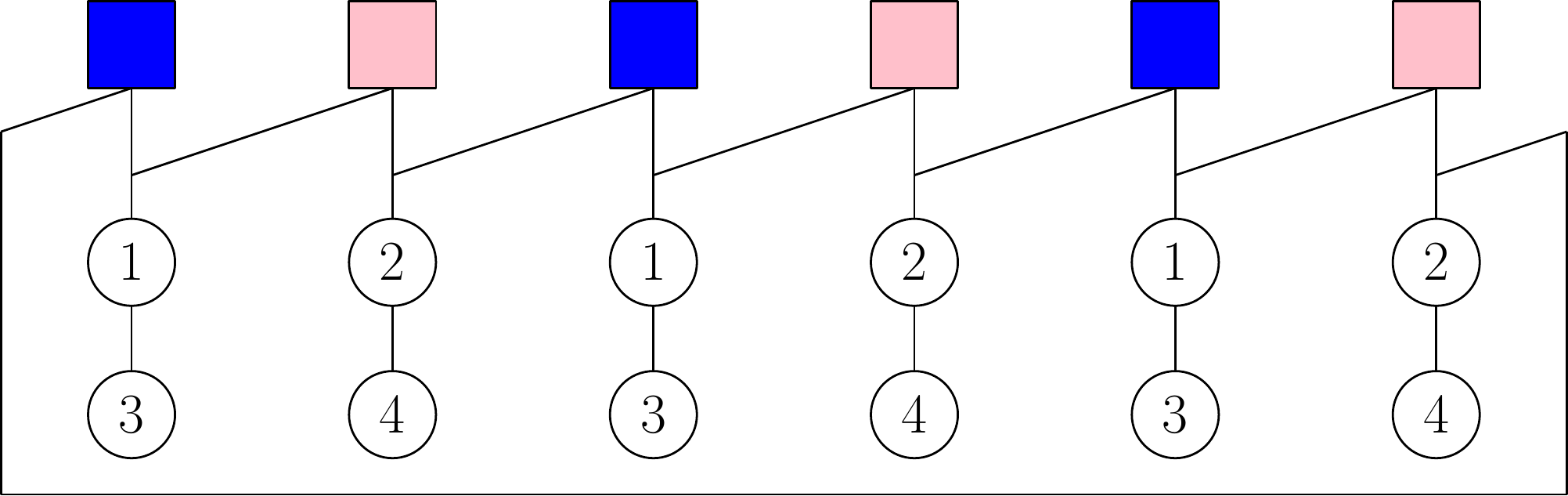}
\caption{A single-level setup with $K=6$, $U=2$, and $d=2$.
The users are divided into $dU=4$ groups (numbered 1 through 4) such that no two users in the same group share any caches.
The caches are colored into $d=2$ colors, such that every user has access to exactly one cache of every color.}
\label{fig:user-division-full}
\end{figure}

%As discussed in Section~\ref{sec:results-achievability},
Coding opportunities are maximized when BS transmissions target users with independent cache contents.
Thus, we begin by dividing the $KU$ users in the system into $dU$ groups, such that no two users in the same group access the same cache (see \figurename~\ref{fig:user-division-full}).
Next, we color the $K$ caches using $d$ colors such that each user has access to exactly one cache of each color.  Finally, we divide every file into $d$ sub-files of equal size, and we color these sub-files using the same $d$ colors used for the caches.
Thus, corresponding to each color, we have $K / d$ caches  and $N$ sub-files of size $F / d$ bits each.

During the placement phase, for each color $c$ we use the random sampling scheme described in Section~\ref{sec:preliminaries} to cache all $N$ $c$-colored sub-files across the $K/d$ $c$-colored caches.
During the delivery phase, we treat each of the $dU$ groups of users and each of the $d$ colors separately.
For every $(\text{user group},\text{color})$ pair, we have a system with $K / d$ users, each with single-access to a cache of memory size $MF$ bits, and $N$ sub-files each of size $F/d$ bits.
Thus, using the rate expression from Lemma~\ref{lemma:single-level-user-access}, the broadcast message size for a particular pair of group and color is given by:
$$
%\left[ \left(\frac{N}{dM}-1\right) \cdot \left( 1 - \left( 1 - \frac{dM}{N} \right)^{K / d} \right) \right]^+ \cdot \frac{F}{d} \mbox{ bits}.  
\left[ \left(N/dM-1\right) \cdot \left( 1 - \left( 1 - dM/N \right)^{K / d} \right) \right]^+ \cdot F/d \mbox{ bits}.  
$$
By repeating the argument for each of the $dU$ user groups and $d$ colors, we get the following result regarding the achievable total BS transmission rate.
\begin{lemma}\label{lemma:single-level}
For $L=1$ level, $K$ caches, $U$ users per cache with an access degree of $d$, $N\ge KU$ files, and memory size $M \in [0, N / d]$, the following BS transmission rate $R^\text{SL}(M,K,N,U,d)$ is achievable:
$$
%\left[ dU \cdot \left(\frac{N}{dM}-1\right) \cdot \left( 1 - \left( 1 - \frac{dM}{N} \right)^{K / d} \right) \right]^+ .
\left[ dU \cdot \left(N/dM-1\right) \cdot \left( 1 - \left( 1 - dM/N \right)^{K / d} \right) \right]^+ .
$$
\end{lemma}

\subsection{Memory-sharing}\label{sec:achievability-memory-sharing}
The  caching-and-delivery scheme described above applies to a single-level system with a given user access structure.
As discussed in Section~\ref{sec:results-achievability}, for a multi-level system, we divide the available cache memory among the various levels and then utilize the above scheme independently for each level.  
What remains is to identify the appropriate choice of the memory-sharing parameters $\{\alpha_i\}_i$ so that the total rate is minimized. 

Clearly, it is desirable to allocate a larger memory to the more popular levels as compared to the less popular ones.
A natural approach would be to use the available memory to completely store the files in the most popular levels, before beginning to cache those in the less popular levels.
Somewhat surprisingly, this approach turns out to be sub-optimal in certain regimes, as illustrated in the following example.

\begin{example}
Let there be $K=8$ caches, and $L=2$ popularity levels such that $N_1=N_2=100$, $(U_1,U_2)=(9,1)$, and $d_1=d_2=1$.
Thus, level 1 is the more popular level, with $U_1/N_1>U_2/N_2$.
Let the $K$ caches have a memory of $M=100$ files.
If we give the whole memory to the more popular level, \emph{i.e.}, $(\alpha_1,\alpha_2)=(1,0)$, then, by \eqref{eq:total-rate} and Lemma~\ref{lemma:single-level}, we can achieve a BS transmission rate of $R=8$.
However, sharing the memory between the levels using $(\alpha_1,\alpha_2)=(\frac32,\frac14)$ yields the smaller rate $R=6$.
Clearly, devoting the entire memory to popular files is sub-optimal for this example setup.
\end{example}

The above example illustrates that a non-trivial choice of memory-sharing parameters $\{\alpha_i\}_i$ might be needed  to achieve the minimum transmission rate for our proposed scheme.
Intuitively, this is due to the diminishing returns property exhibited by the rate-memory tradeoff for each individual level.
With that in mind, we now describe our choice for these parameters. 
We begin by partitioning the set of levels into three subsets, denoted by $H$, $I$, and $J$, as follows.
\begin{definition}[$M$-feasible partition]\label{def:m-feasible}
For any cache memory $M$, we define an \emph{$M$-feasible partition} $(H,I,J)$ of the set of levels $\{1,\ldots,L\}$ as one that satisfies:
\begin{IEEEeqnarray*}{lCl'rCcCl}
\forall h &\in& H, & && \tilde M &<& (1/K)\sqrt{N_h/U_h} + y_h;\\
\forall i &\in& I, & (1/K)\sqrt{N_i/U_i} &\le& \tilde M &\le& (1/d_i)\sqrt{N_i/U_i};\\
\forall j &\in& J, & (1/d_j)\sqrt{N_j/U_j} &<& \tilde M,
\end{IEEEeqnarray*}
where
\(
\tilde M = \frac{ M - \sum_{j\in J} N_j / d_j }{ \sum_{i\in I}\sqrt{N_iU_i} }
\)
and
\(
y_h = \frac{N_h/K}{\sum_{i\in I}\sqrt{N_iU_i}}
\).
\end{definition}

We now describe how to assign $\alpha_i$'s for the subsets $H$, $I$, and $J$.
The set $H$ will consist of levels that get zero cache memory, \emph{i.e.}, $\alpha_hM=0$ for all $h\in H$.
In contrast, the set $J$ will consist of levels that get enough cache memory so that all user requests for these levels will be satisfied without any broadcast.
In other words, for all $j\in J$, we set $\alpha_jM= N_j / d_j$ so that the individual rate $R_j=0$ for level $j$, as determined by Lemma~\ref{lemma:single-level}.
Finally, the set $I$ will consist of the remaining levels, which will share any remaining cache memory $\left(M-\sum_{j\in J} N_j / d_j\right)$ according to their relative popularity.
We formalize this memory allocation in the following definition.
\begin{definition}[Popularity-aware memory allocation (PAMA)]\label{def:poma}
For an $M$-feasible partition $(H,I,J)$ of the set of levels $\{1,\ldots,L\}$, the PAMA $(\alpha_1,\ldots,\alpha_L)$ is:
\begin{IEEEeqnarray*}{lCl"rCl}
\forall h &\in& H, & \alpha_hM &=& 0;\\
\forall i &\in& I, & \alpha_iM &=& \frac{\sqrt{N_iU_i}}{\sum_{i'\in I}\sqrt{N_{i'}U_{i'}}} \left( M - \sum_{j\in J}\frac{N_j}{d_j} \right);\\
\forall j &\in& J, & \alpha_jM &=& \frac{N_j}{d_j}.
\end{IEEEeqnarray*}
\end{definition}

It can be easily shown that $\alpha_i\in[0,1]$ for all $i$. and $\sum_{i=1}^L\alpha_i=1$.
Note that for a level $i \in I$, the amount of (remaining) memory given \emph{per file} is proportional to $\sqrt{U_i / N_i}$, which is a measure of the popularity of $i$.

For any memory size $M$, we choose the memory-sharing parameters $\{\alpha_i\}_i$ to be any PAMA corresponding to $M$.
With these parameters, the total achievable BS transmission rate for the scheme can now be evaluated using Lemma~\ref{lemma:single-level} for each level, and immediately gives us the result in Theorem~\ref{thm:multi-level-achievability}.

Note that, given all the problem parameters and a memory size $M$, it is possible to numerically compute the optimal values of $\alpha_i$ (using any convex optimization solver).
However, our analysis above provides a structured and exact solution, the advantages of which are two-fold.
First, this enables us to design an \emph{efficient algorithm}\myextendedonly{\footnote{The algorithm runs in $\Theta(L^2)$ time.}} that finds a PAMA \remove{optimal memory allocation} \emph{for all values of $M$}, the details of which can be found in \myextendedref{app:achievability}.
Secondly, the structure of the solution, particularly with regards to the partitioning of the set of levels into $(H,I,J)$, is what enables the analysis of the gap between the achievable rate and the information-theoretic lower bounds, and ultimately allows us to demonstrate the order-optimality of ML-PAMA.

\subsection{Extensions}

The scheme described so far is designed for the very symmetric and regular setup in Section~\ref{sec:model}.
While the design is based on several assumptions, we here explore extensions for when the following two assumptions no longer hold: \textsf{(1)} Equal file sizes; and \textsf{(2)} Linear arrangement of the APs.

Firstly, in practice, files at a certain host server do not have the exact same size.
Sizes of video files, for instance, depend on the length and quality of the video.
For such a scenario, we propose the following modification to the scheme.
%Recall that, in the simplest settings, our scheme currently samples a random subset of size $\frac{MF}{N}$ bits from each file, totaling $MF$ bits.
Recall from Example~\ref{ex:coded-caching} that, when all files have equal size $F$, each cache samples a random subset of size $\frac{MF}{N}$ bits from each file, totaling $MF$ bits.
In effect, we are taking a random subset of size $MF$ bits out of the total $NF$ bits available at the server.
If files had different sizes $F_1,\ldots,F_N$, and $F$ was their average, then we can achieve a similar result by caching a random subset of size $MF$ bits, out of the total $\sum_{i=1}^N F_i$ bits.

Secondly, while the linear arrangement of APs was convenient for the analysis of the lower bounds, a two-dimensional arrangement fits a realistic setting more closely.
A natural extension of the model would be to place APs on a 2D lattice.
We can then color the caches using $d$ colors such that the discrete Voronoi region of any point in the plane will contain at least one cache of each color.
From this point on, the scheme is exactly the same.

\section{Information-theoretic impossibility results}\label{sec:converse}

Theorem~\ref{thm:multi-level-optimality} establishes the order-optimality of ML-PAMA, for any multi-level system, by comparing its achievable BS transmission rate with the information-theoretic optimum.

To prove this result, we derive lower bounds on the rate of \emph{any} feasible scheme, as a function of the problem parameters $K$, $L$, $\{N_i, U_i, d_i\}$ and memory size $M$.
In particular, we provide two types of lower bounds.
The first are cut-set (or cooperative) bounds\myextendedonly{ \cite{coverbook}}, which are based on the idea that, if one considers a subset of users and artificially allows them to cooperate with each other by sharing the content of the caches they access, then the transmission rate required will be a lower bound on the rate for the original problem setting (with no cooperation).
The resulting bound is given as follows.
\begin{lemma}[Cut-set bounds]\label{lemma:converse-cutset}
For any level $i\in\{1,\ldots,L\}$ and any $v\in\{1,\ldots,KU_i\}$, the optimal rate is bounded by:
\[
R^\ast(M) \ge v - \frac{ \ceil{v/U_i} + (d_i-1) }{ \floor{ N_i/v } }M.
\]
\end{lemma}

Note that the cut-set bound only considers one popularity level at a time and assumes that only user requests of this level need to be served.
While such cut-set bounds are enough to prove order-optimality for a single-level system \cite{maddah-ali2012, maddah-ali2013}, they do not suffice for a multi-level system.
Hence one needs to derive more general lower bounds that take into account the constraint that both the available resources need to serve requests for files from multiple levels.
%Hence one needs to derive more general lower bounds that take into account the constraint that both the available memory as well as the BS transmission need to serve requests for files from multiple levels.
The fact that the different levels have different access structures associated with them makes it even more challenging to derive such bounds for our setup.
Below, we present one such class of non-cut set lower bounds. 
\begin{lemma}[Non-cut set bounds]\label{lemma:converse-general}
For any $A\subseteq\{1,\ldots,L\}$, any $l\in\{1,\ldots,L\}$ with $l\notin A$, any $s\in\{d_l,\ldots,     K\}$, and any $b\in\mathbb{N}^+$, we have:
\begin{IEEEeqnarray*}{rCl}
R^\ast(M) &\ge& \frac{1}{D+1} \min\left\{ (s-d_l+1)U_l \,,\, \frac{N_l}{sb} \right\}\\
&& {} + \sum_{j\in A} \min\left\{ U_j \,,\, \frac{N_j}{bd_j} \right\} - \frac{M}{b}.
\end{IEEEeqnarray*}
\end{lemma}

Both lemmas are proved in \myextendedref{app:converse-proof}.

Next, we analyze the gap between these lower bounds and the achievable rate of ML-PAMA.
Recall that the scheme divides the popularity levels into a partition $(H,I,J)$.
We go over all feasible $(H, I, J)$ partitions that can arise for different multi-level systems, and for each of them carefully choose the parameters in the above lemmas to get the best possible lower bound.
We then analyze the gap between this lower bound and the achievable rate associated with the $(H, I, J)$ partition under consideration.
We are able to show that, for any multi-level system, the achievable rate and the information-theoretic lower bound differ by at most a constant multiplicative gap.
This gives us the order-optimality result in Theorem~\ref{thm:multi-level-optimality}.
Details of the gap analysis can be found in \myextendedref{app:gap}.

\section{Discussion and numerical evaluations}
\label{sec:discussion}\label{sec:numerics}

In the previous sections, we presented theoretical results for any given set of popularity levels and associated user access structures.
However, in practice, what is available is a ``continuous'' popularity distribution over the entire set of files, and it is up to the designer to choose: \textsf{(a)} the number of popularity levels; \textsf{(b)} which files to assign to which level; and \textsf{(c)} the corresponding user access degree for each popularity level.
For each such choice, our theoretical results characterize the minimum broadcast transmission rate, and we study, in this section, the impact of these choices on the transmission rate.
Furthermore, while our theoretical model assumed that, for each popularity level, the number of users per AP is exactly the same, we relax this assumption here by allowing each user to randomly connect to one of the $K$ APs and request a file stochastically, according to the underlying popularity distribution.
Finally, we will also compare the performance of our scheme with that of the traditional LFU approach, as well as the information-theoretic lower bounds presented before.

We use a YouTube dataset \cite{YoutubeRepository} for our evaluations, which provides the number of requests for around $500\,000$ videos.
\figurename~\ref{fig:youtube} shows the popularity distribution of the files, which resembles a Zipf distribution similar to those commonly observed for multimedia content\myextendedonly{ \cite{breslau1999web}}. 

\begin{figure}
\centering
\includegraphics[width=\myplotwidth]{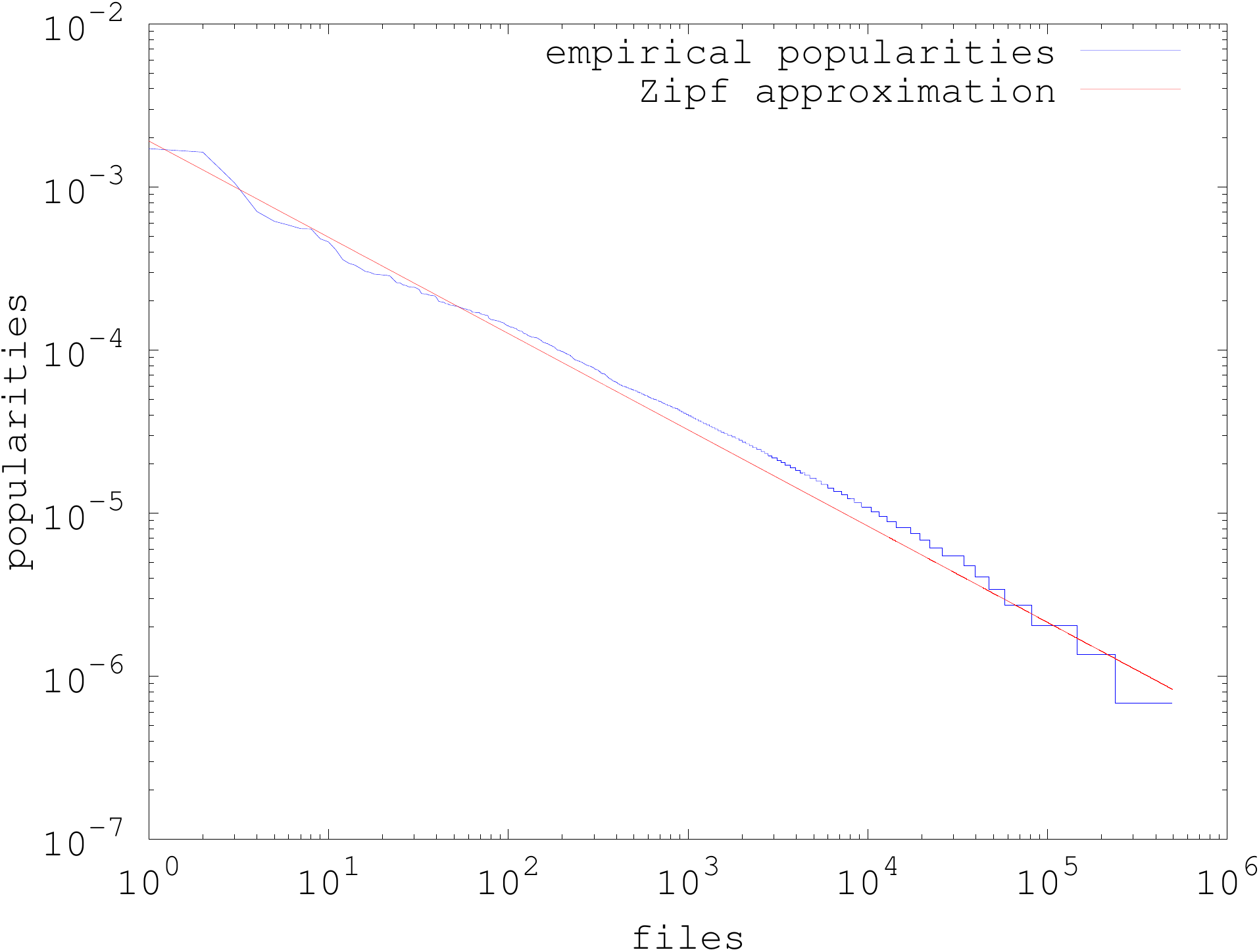}
\caption{Popularity of $493\,722$ YouTube videos.
The distribution can be approximated by a Zipf distribution with parameter $0.6$.}
\label{fig:youtube}
\end{figure}

\subsection{Discretizing a continuous popularity distribution}

\begin{figure}
\centering
\includegraphics[width=\myplotwidth]{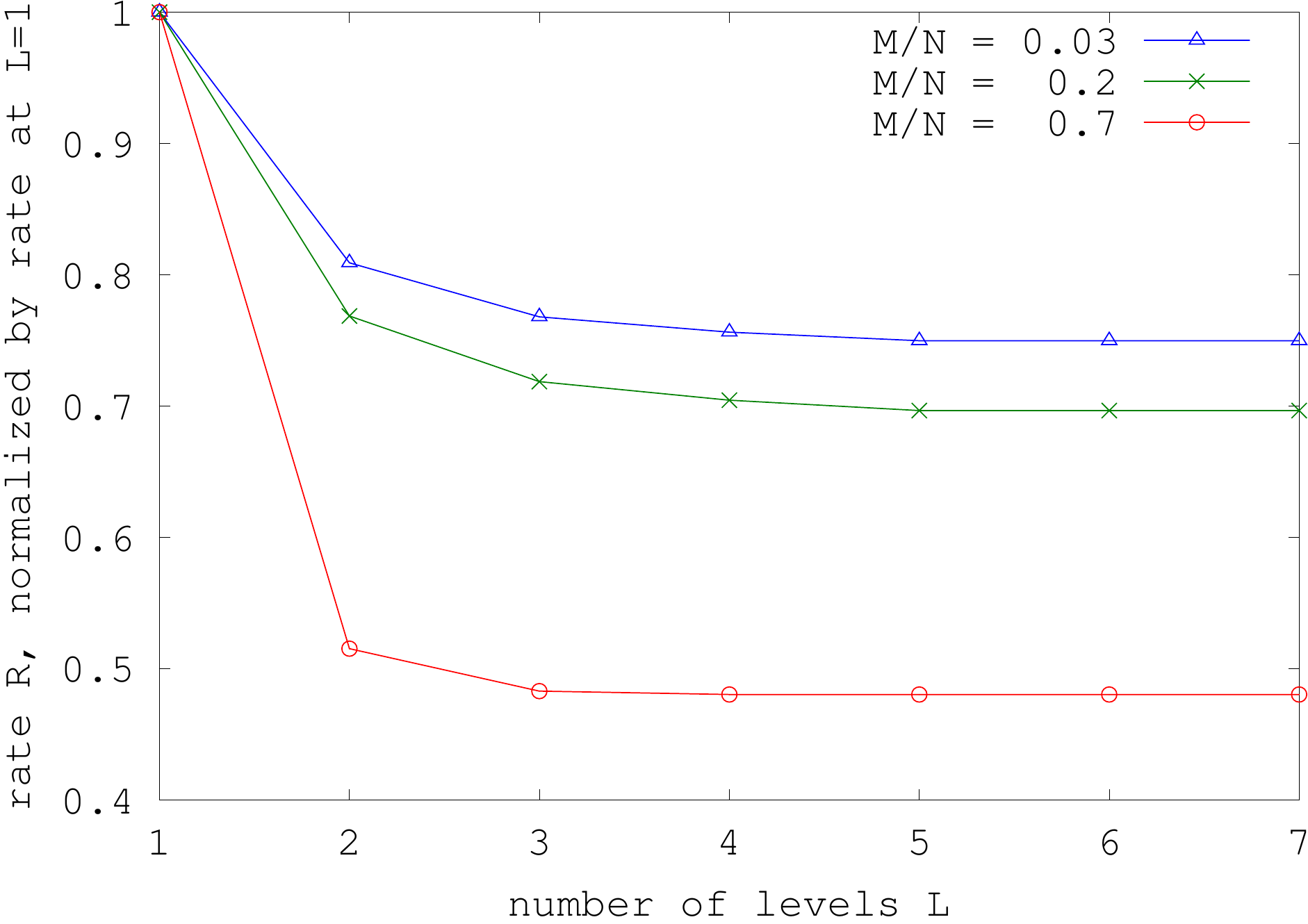}
\caption{Rate achieved by \myreplace{the memory-sharing scheme}{ML-PAMA} vs.\ number of levels, for different values of cache memory.
For each $L$, we choose the $L$ levels that minimize the achievable rate (using brute-force search).
\myinsert{For ease of comparison,} the rate values \myreplace{are}{have been} normalized by the rate at $L=1$.}
\label{fig:youtube-R-vs-L}
\end{figure}

Our first step is to divide the files in the YouTube dataset into a certain number of levels, based on the popularity profile in \figurename~\ref{fig:youtube}.
We consider small, moderate, and large values of $M / N$ ($0.03$, $0.2$, and $0.7$) and set the user access degree $d_i = 1$ for every level $i$, so as to study the impact of the number of levels on the broadcast rate in isolation.
For increasing values of $L$, we find the division of the files into $L$ levels that minimizes the rate achieved by the ML-PAMA scheme using a brute-force search.
We plot the minimum achievable rate versus $L$
in \figurename~\ref{fig:youtube-R-vs-L}.
As is easily apparent, while there is a significant gain in performance between treating all files as one level and dividing them into two levels, the gain    decreases with diminishing returns as $L$ increases.
This shows the importance of dividing files into multiple levels, but also suggests that 3--4 levels are sufficient to derive most of the benefits.

\begin{figure}
\centering
\includegraphics[width=\myplotwidth]{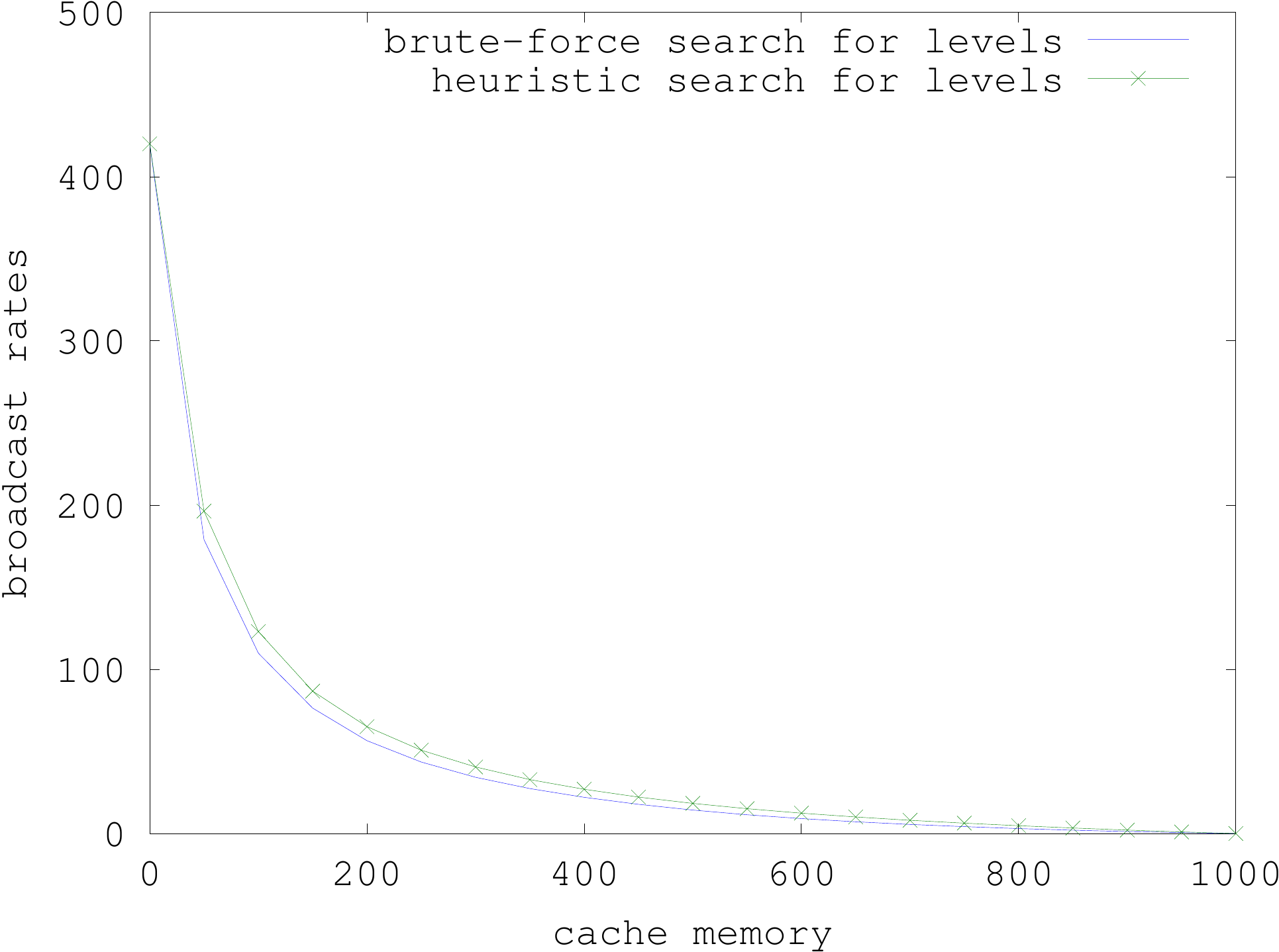}
\caption{Performance of the heuristic algorithm \mydelete{as }compared with \myinsert{that of} the brute-force (optimal) search over two levels, when $N=10\,000$ files follow a $\mathrm{Zipf}(0.6)$.
The maximum multiplicative loss \myreplace{is $\le1.92$}{does not exceed $1.92$}.}
\label{fig:heuristic-vs-brute-force}
\end{figure}

While the brute-force search described above yields the optimal partition for any given number of levels $L$, it is computationally expensive and has a running time of at least $\Omega(N^{L-1})$.
We \myreplace{describe}{design}\mycomment{ (we don't \emph{describe} it in the short version)} a heuristic \myreplace{scheme}{algorithm} for dividing the content into two levels, \mydelete{which is }based on analyzing the broadcast rate for any given partition when the underlying popularity profile is a true Zipf distribution.
We skip it here for lack of space and provide it in \myshortonly{\cite{extended}}\myextendedonly{Algorithm~\ref{alg:simple-split-2}}.
The heuristic \myreplace{scheme}{algorithm} \myreplace{is much faster than the brute-force search}{runs in constant time} and, as shown in \figurename~\ref{fig:heuristic-vs-brute-force}, performs almost as well as the \myreplace{latter}{brute-force search}.
As a result, we use \myshortonly{this heuristic}\myextendedonly{Algorithm~\ref{alg:simple-split-2}} in all our experiments to discretize the popularities.
\mydelete{In particular, it will, for most values of $M$, divide the $500\,000$ YouTube files into $(N_1,N_2)\approx(150\,000,350\,000)$.}

\begin{comment}
\begin{figure*}
\centering
\subfloat[Zipf distribution with parameter $s=0.6$]{
\includegraphics[width=\myplotwidth]{h-vs-bf-06.pdf}
\label{fig:h-vs-bf-06}
}
\hfil
\subfloat[Zipf distribution with parameter $s=1.5$]{
\includegraphics[width=\myplotwidth]{h-vs-bf-15.pdf}
\label{fig:h-vs-bf-15}
}
\caption{Performance of the heuristic algorithm as compared with the brute-force (optimal) search over two levels.
Two cases are considered, one (a) where the Zipf parameter is $0.6$, and the other (b) when it is $1.5$.
The maximum multiplicative loss is $\le1.92$ in (a) and $\le3.44$ in (b).}
\label{fig:heuristic-vs-brute-force}
\end{figure*}
\end{comment}

\myextendedonly{
\begin{algorithm}
\caption{\small Two-level splitting algorithm on Zipf distributions}
\label{alg:simple-split-2}
\begin{algorithmic}[1]
\small
\Procedure{ZipfSplit}{$s,N,K,M$}

\If{$s<1$}
  \State $n \gets \frac{1-s}{2-s} \cdot \min\left\{ MK , N \right\}$
\Else
%  \State Let $m_1$ and $m_2$ be:
%  \begin{IEEEeqnarray*}{rCl}
%  m_1 &=& \min\left\{ \frac{N}{K} , K^{\frac{1}{s-1}} \right\}\\
%  m_2 &=& \max\left\{ N^{\frac{1}{s}} , K^{\frac{1}{s-1}} \right\}
%  \end{IEEEeqnarray*}
  \State $m_1 \gets \min\left\{ N/K , K^{\frac{1}{s-1}} \right\}$
  \State $m_2 \gets \max\left\{ N^{\frac{1}{s}} , K^{\frac{1}{s-1}} \right\}$
  \If{$M \le m_1$}
	\State $n\gets\left(MK\right)^{1/s}$
  \ElsIf{$m_1 < M < m_2$}
	\State $n\gets N^{1/s}$
  \ElsIf{$M \ge m_2$}
	\State $n\gets 0.1 M$
  \EndIf
\EndIf
\State $\left(N_1,N_2\right) \gets \left(n,N-n\right)$
\State \Return $(N_1,N_2)$
\EndProcedure

\end{algorithmic}
\end{algorithm}
}

\subsection{Impact of multi-access on the achievable rate}
\begin{figure}
\centering
\includegraphics[width=\myplotwidth]{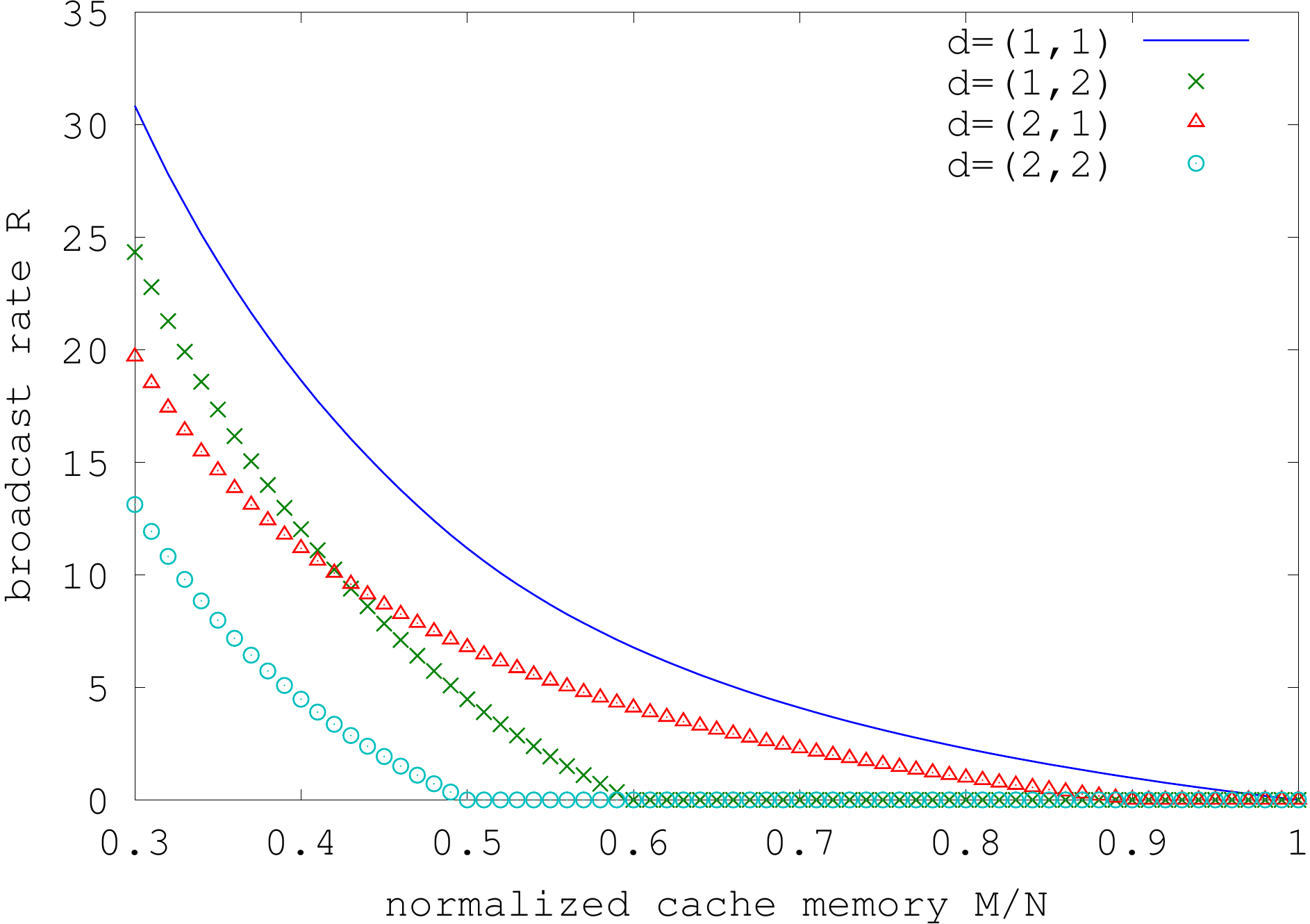}
\caption{Achievable rate vs.\ cache memory in a two-level setup, for different access structures.}
\label{fig:youtube-multi-access}
\end{figure}

To study the effect of multi-access in isolation, we will fix the partition we use to divide the files into different levels and then look at different multi-access structures.
Consider $L = 2$ levels, with $N_1 = 0.2N$, and $N_2 = 0.8N$ files.
We plot in \figurename~\ref{fig:youtube-multi-access} the BS rate of our scheme as a function of the normalized memory $M / N$, for four different access structures $(d_1, d_2)$: $(1,1)$, $(1, 2)$, $(2,1)$, and $(2,2)$. 
%For \figurename~\ref{fig:youtube-multi-access}, we consider $L = 2$ levels with $N_1 = 0.2N$, $N_2 = 0.8N$ files respectively, and then plot the BS rate of our poposed scheme as a function of the normalized memory $M / N$, for four different access structures $(d_1, d_2)$: $(1,1)$, $(1, 2)$, $(2,1)$, and $(2,2)$. 

As one would expect, allowing for multi-access greatly improves the transmission rate. %the rate decreases as we give greater cache access to users.
For example, the rate for the multi-access system with $(d_1 = 2, d_2 = 2)$ is smaller than the rate for the single-access system with $(d_1 = 1, d_2 = 1)$.
The cases $(d_1 = 1, d_2 = 2)$ and $(d_1 = 2, d_2 = 1)$ provide a more interesting comparison.
For small memory size $M$, the former gives a lower rate since the cache memory mainly contains files from level $1$, and so giving higher access to level $1$ is more beneficial in reducing the rate.
On the other hand, as $M$ grows and files from level $2$ start occupying a significant portion of the memory, it becomes more efficient to give higher access to level $2$ since it has many more files than level $1$. 

\begin{figure}[t]
\centering
\includegraphics[width=\myplotwidth]{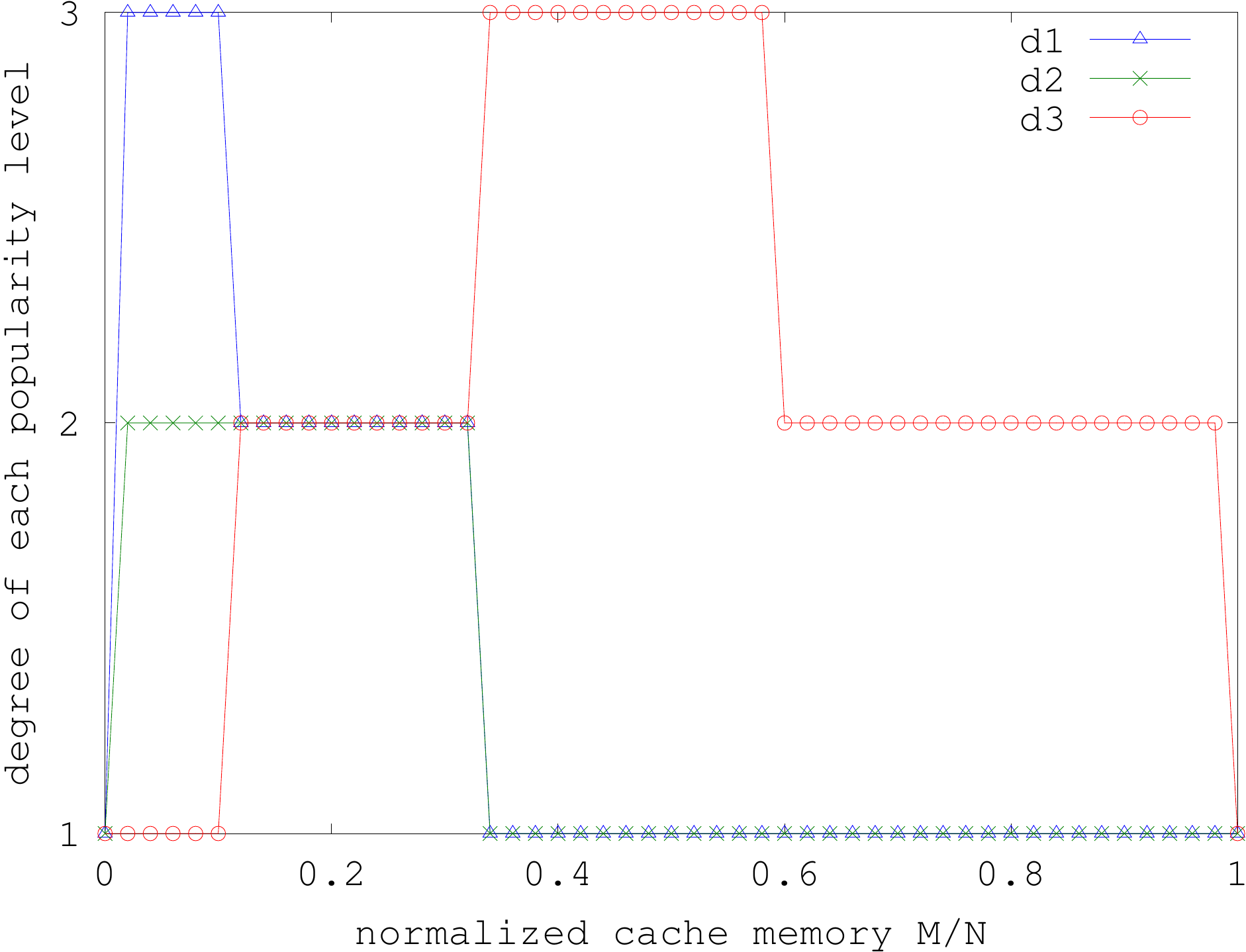}
\caption{Optimal access structure vs.\ memory, with $d_{\max}=3$, $d_\mathrm{avg}=2$.}
\label{fig:youtube-dvM}
\end{figure}

While greater cache access helps reduce the rate, there is also a cost associated with it in terms of the increased delay in gathering data from multiple APs, as well as a reduced rate as a user connects to farther APs.
In general, for a given multi-level setup with parameters $L$, $K$, $\{N_i,U_i\}$, and $M$, such a cost can be included in the rate optimization framework
as one or more inequalities of the form $\mathrm{cost}_j(K,\{U_i,d_i\}_i)\le C_j$,
% as follows:
%\begin{IEEEeqnarray*}{c"c}
%\underset{(d_1,\ldots,d_L)}{\text{minimize}} & R\left(M,K,\{N_i,U_i,d_i\}_i\right)\\
%\text{subject to} & \mathrm{cost}(K,\{U_i,d_i\}_i) \le C,
%\end{IEEEeqnarray*}
for some maximum cost $C_j$.
%where $\mathrm{cost}_j(\cdot)$ is some cost function relating to the access structure, and $C_j$ is the maximum allowed cost.
The above optimization problem can be numerically solved by a designer in order to identify the optimal access structure for the multi-level system under consideration.
However, to derive some intuition about how the costs impact the optimal multi-access structure, let us consider a setup with $L = 3$ levels, and with $N_1 = 0.04N$, $N_2 = 0.13N$, and $N_3 = 0.83N$ files in the three levels.
Say we want to include both a maximum degree constraint $d_i \le 3$ for each level $i$, as well as an average degree constraint $(\sum_i U_i d_i ) / U \le 2$.
Then, \figurename~\ref{fig:youtube-dvM} plots the optimal access structure vs.\ the normalized memory size.
As before, when the memory is small, the optimal access structure is one which satisfies $d_1\ge d_2\ge d_3$, but this relation becomes reversed as the memory increases.

\subsection{Stochastic variations in user profiles}
The theoretical setup and results presented in the previous sections assumed a symmetric and deterministic user profile across all the APs.
In particular, exactly $U_i$ users are assigned to each AP to request files from level $i$.
This section aims at evaluating the robustness of ML-PAMA to asymmetry and stochasticity in the user profiles across caches.

\begin{figure}
\centering
\includegraphics[width=\myplotwidth]{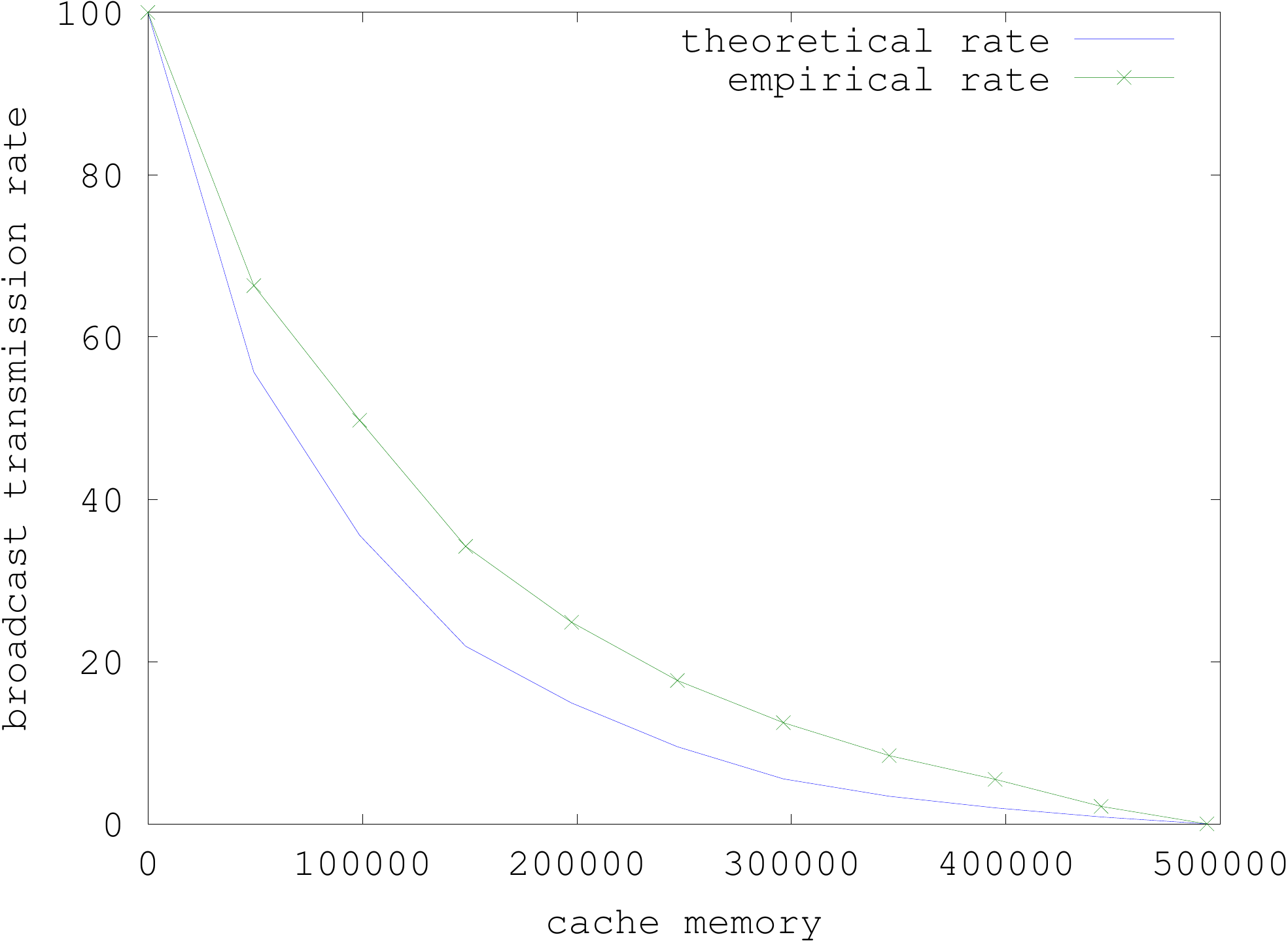}
\caption{Comparison of the theoretical rate with the empirical rate, based on simulations of demands over the YouTube dataset, with $5$ caches and $100$ total users.
The theoretical rate is off by a factor of up to $2.8$ from the empirical.}
\label{fig:simulations}
\end{figure}

We consider a setup where each of the $KU$ users in the system randomly connects to one of the $K$ APs and requests a file stochastically, according to the YouTube popularity distribution in \figurename~\ref{fig:youtube}.
We ran simulations for this setup, and we plot the empirically achieved rate against the cache memory in \figurename~\ref{fig:simulations}.
For comparison, we also show the rate predicted by our theoretical model, which assumes a symmetric user profile across the caches.
Clearly, the theory very closely predicts the empirical results for a random user profile, thus demonstrating the robustness of our theoretical results to stochastic variations across APs and justifying their utility in practice.  

%\myextendedonly{
\subsection{Comparison with \myreplace{LFU}{Least-Frequently Used (LFU)}}

\mycomment{I felt this part could use some improvement.}

\begin{figure}
\centering
\includegraphics[width=\myplotwidth]{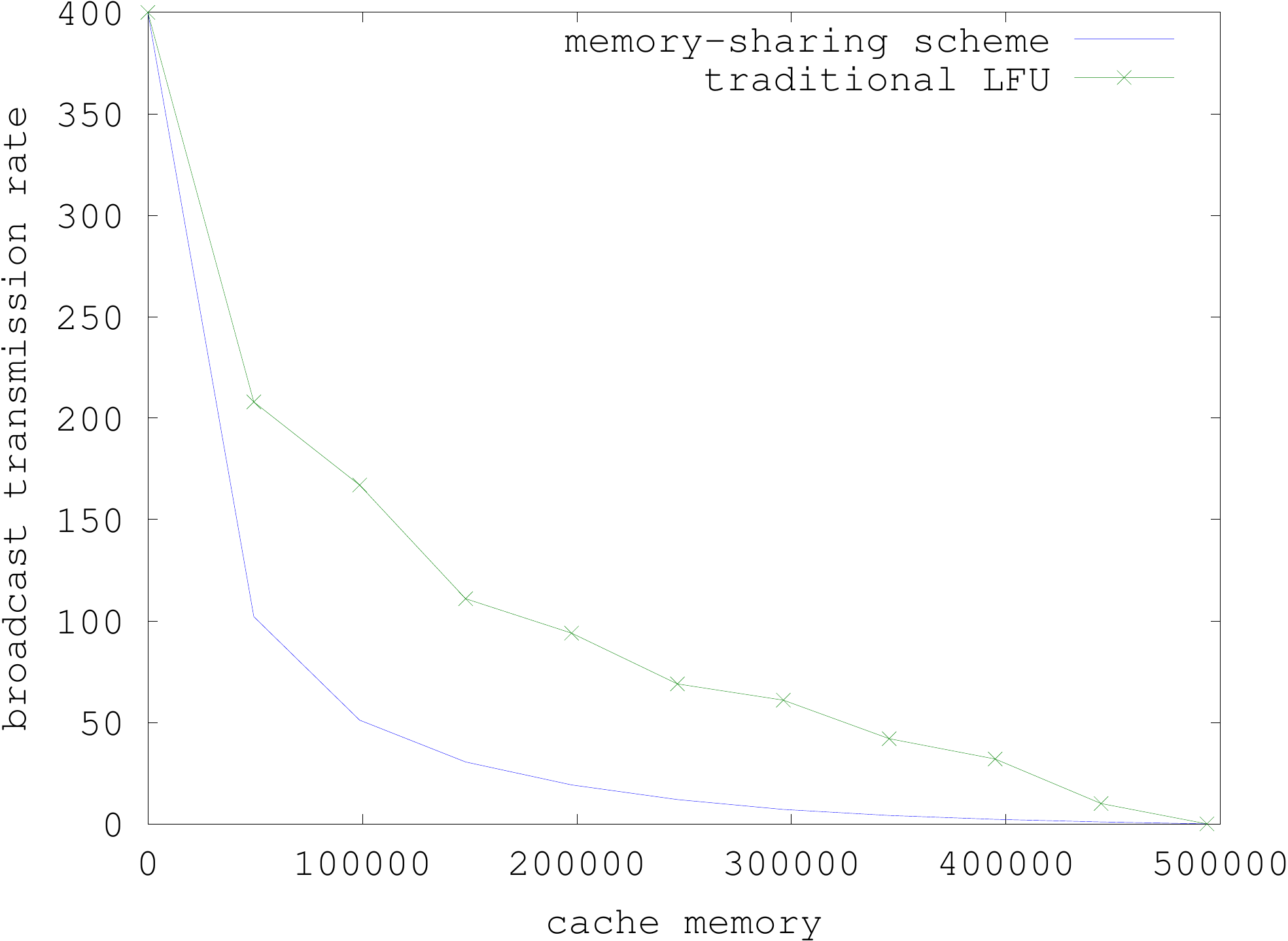}
\caption{Comparison of the memory-sharing scheme \myinsert{(ML-PAMA)} with traditional LFU.
\myreplace{The memory-sharing scheme}{ML-PAMA} achieves up to a factor-$14.5$ in gain over LFU.}
\label{fig:lfu-comparison}
\end{figure}

\mydelete{
In \figurename~}\mydeletem{\ref{fig:lfu-comparison}}\myreplace{, we compare our scheme with the traditional Least-Frequently-Used (LFU) scheme.
The LFU fully stores the $M$ most popular files, such that requests for more popular files are completely served from the cache, and requests for less popular files must be directly handled by the BS transmission.
}
{
In this section, we compare the performance of ML-PAMA with that of the traditional LFU scheme using simulations on the YouTube data.
For any memory size $M$, LFU fully stores the $M$ most popular files, so that requests for more popular files are completely served from the cache, and requests for less popular files are fully handled by the BS transmission.
The results, given in \figurename~\ref{fig:lfu-comparison}, show the superiority of ML-PAMA.
}

%}

\remove{
In \figurename~\ref{fig:lfu-comparison}, we compare our scheme with the traditional Least-Frequently-Used (LFU) scheme, as well as the scheme for coded caching with non-uniform demands presented in \cite{niesen2013}.
The LFU scheme fully stores the $M$ most popular files, such that requests for more popular files are completely served from the cache, and requests for less popular files must be handled by the BS transmission.
The scheme in \cite{niesen2013}, on the other hand, is similar to ours, but with two differences.
The first is that popularity levels are created such that the popularities of files in the same level differ by at most a factor of two.
The second is that all the $L$ levels are given equal cache memory, \emph{i.e.}, $\alpha_i=1/L$ for all levels $i$.
For this reason, we refer to it as the equal-partitioning scheme.
}

\subsection{Numerical gap}\label{sec:numerics-gap}

\myshortonly{

As discussed in Section~\ref{sec:results}, the multiplicative gap in Theorem~\ref{thm:multi-level-optimality} results from many over-simplifications in bounding the achievable rate.
Numerical results suggest that this gap is in fact much smaller.
For example, consider the setup with $L=3$, $K=10$, $(N_1,N_2,N_3)=(500,1500,8000)$, $(U_1,U_2,U_3)=(9,5,1)$, and $(d_1,d_2,d_3)=(1,3,5)$.
Then the maximum gap between the achievable rate and the lower bounds is approximately $6$, which is
%Even in the worst case, the gap is still
much smaller than predicted by the conservative theoretical bounds.
In the worst case, it does not exceed $45$ when $L=3$ and $d_i\le5$.

}

\myextendedonly{
As discussed in Section~\ref{sec:results}, the multiplicative gap in Theorem~\ref{thm:multi-level-optimality} results from many over-simplifications in bounding the achievable rate.
Numerical results suggest that this gap is much smaller.
For instance, when $D=5$ and $L=3$, the worst-case gap does not exceed $45$.
We provide an example in \figurename~\ref{fig:gap}, where the gap between the achievable rate and the lower bounds is approximately $6$, which is much smaller than predicted by the conservative theoretical bounds.

\begin{figure}
\centering
\includegraphics[width=\myplotwidth]{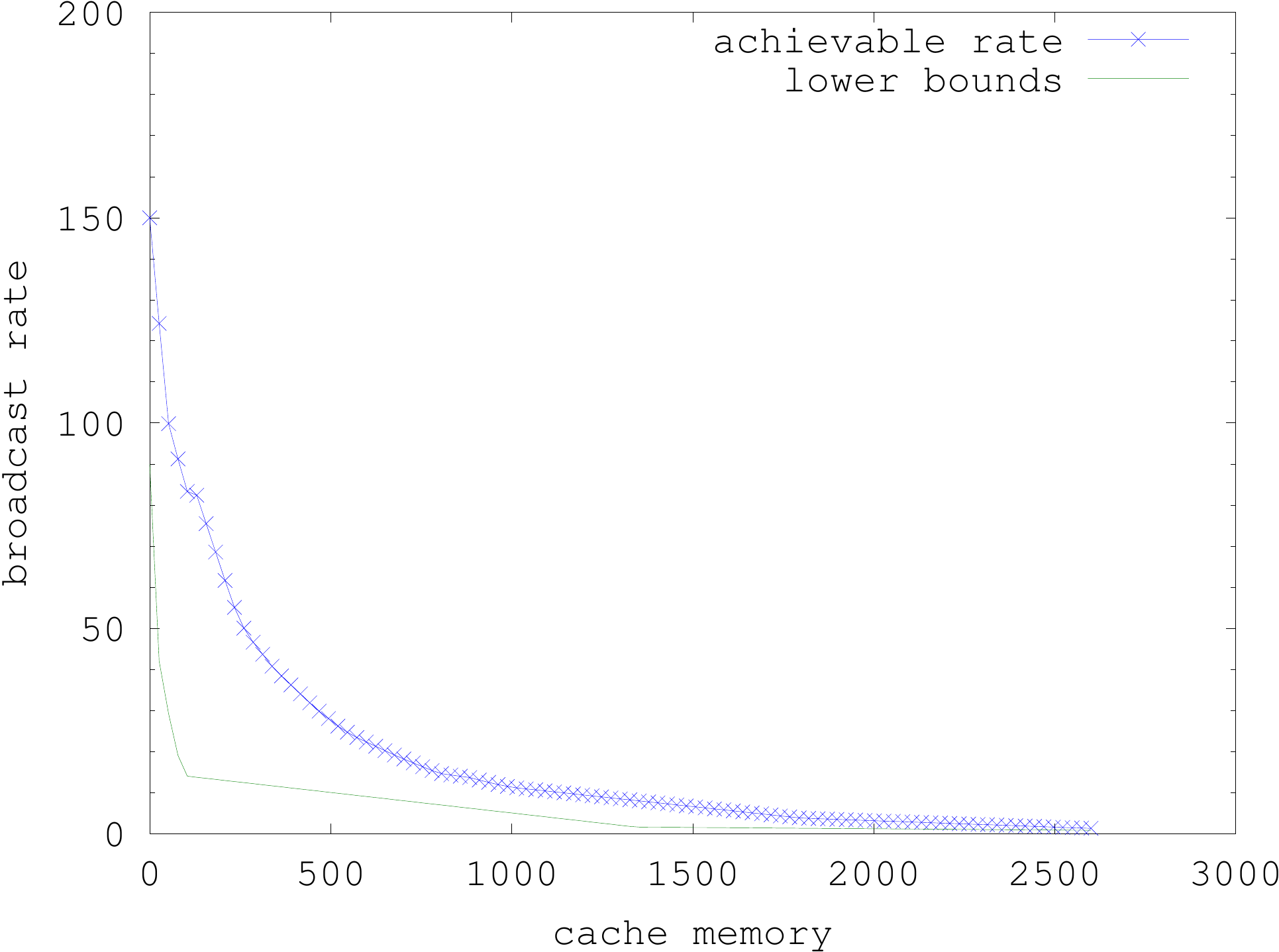}
\caption{Rate achieved by the memory-sharing scheme compared with the information theoretic lower bounds.
The setting is as follows: $L=3$, $(N_1,N_2,N_3)=(500,1500,8000)$, $K=10$, $(U_1,U_2,U_3)=(9,5,1)$, and $(d_1,d_2,d_3)=(1,3,5)$.
The maximum gap is $\approx6$.}
\label{fig:gap}
\end{figure}
}

\section{Related work}\label{sec:related}
Content caching-and-delivery has an extensive history and has been widely studied; see for example \cite{Wessels:2001} and references therein.
More recently, it has
been studied in the context of Video-on-Demand systems, for which
efficient content placement and delivery schemes have been proposed in
\cite{Borst:2010, Tan:2013, llorca2013}.
The impact of different popularities
among the available content on the caching schemes has also been
investigated; see for example \myshortonly{\cite{Applegate:2010}}\myextendedonly{\cite{breslau1999web,
Applegate:2010, jin2000popularity}}.
A common feature among the
conventional caching schemes studied in the above literature
is that
the stored content is replicated across the caches in the system.
When users request content, the parts of the
demanded files that are available at nearby caches are served
locally.
The remaining file parts are served via orthogonal unicast
transmissions by a central server hosting all the files.

The inherent broadcast nature of wireless communications presents an opportunity to improve system performance by serving multiple users simultaneously.
Recently, \cite{maddah-ali2012, maddah-ali2013} proposed new caching
schemes for a single-hop broadcast network, which, unlike the
conventional schemes, are based on storing different data across the
caches in the network during placement and then using coded-multicast
server transmissions during delivery to satisfy multiple user requests
simultaneously.
These observations were extended to the case of two-tier tree networks in \cite{Hcodedcaching} and to non-uniform file popularities in \cite{niesen2013, MLcodedcaching, Zcodedcaching}.
Our work differs from all the above in that we utilize the ability of users to access multiple AP caches to design a system architecture that dynamically allocates user access to APs depending on the requests, significantly improving performance.
The main technical difficulty here is the partial overlap between the caches accessed by various users in the system: both the achievability scheme as well as the lower bounds require new ideas to take this overlap into account.
For any access profile, we are able to prove the order-optimality of our scheme by comparing its achievable rate to information-theoretic lower bounds.
We also numerically evaluate the scheme's performance over a YouTube dataset and demonstrate its benefits over single-AP access.
This evaluation further enables us to study the optimal access profile for different values of memory, as well as optimize the number of popularity levels to demonstrate that 3--4 levels are sufficient to accumulate most of the gains of memory-sharing. 

In terms of the system architecture, our vision is closest to
\cite{FemtoCaching, SCCaching, HCNcaching}, which recently
proposed a caching architecture for heterogeneous wireless networks,
with the small-cell or WiFi access points acting as helpers by storing
part of the content.  However, they did not utilize the broadcast
property enabling network coded delivery,
%Moreover, they did not
nor did they
examine the problem from an information-theoretic viewpoint,
\emph{i.e.}, develop impossibility results that are not tied to a
particular scheme.

There has also been a lot of work in algorithms and protocols related
to content distribution over networks.
Content distribution networks
such as those managed by Akamai and Amazon are integral to how content
is served over the internet today, and there has been a lot of work on
their system design and
architecture \myshortonly{\cite{nygren2010akamai}}\myextendedonly{\cite{nygren2010akamai,pathan2007taxonomy}}.
Over the last few years, a new paradigm for how
internet protocols should search for and deliver content, called
\emph{content-centric networking}, has emerged that has garnered
significant attention; see for example  \cite{ahlgren2012survey,
Fayazbakhsh:2013, psaras2012probabilistic}  and references therein.
These works have been mainly from a networking
systems viewpoint and a significant focus has been on naming data
objects and routing for content delivery.
In this work, we study such content-centric networks from an information-theoretic viewpoint, and our focus is on the last-hop wireless link.

\myextendedonly{
Various other methods for content
distribution have been proposed in the literature, for example
\cite{Ioannidis:2010} explores
distributed caching in mobile networks using device-to-device
communications and \cite{li2011codeon} studies content distribution for vehicular  networks.
Other related work includes \cite{Gitzenis13} which derives scaling laws for content replication in multihop wireless networks; \cite{AltmanAG13} which studies the benefit of coded caching when the caches are distributed randomly; and \cite{YangH13} which explores the benefits of adaptive content placement, using knowledge of user requests.
However, these have not been examined from an
information-theoretic viewpoint and the focus has been on optimizing particular strategies.
}

%%fakesection{bibliography}
\bibliographystyle{IEEEtran}
\myextendedonly{\bibliography{caching}}
\myshortonly{\bibliography{caching-abbrv}}

\appendices
\section{Proof of the order-optimality of ML-PAMA (Theorem~\ref{thm:multi-level-optimality})}\label{app:order-optimal-proof}\label{app:gap}

This section aims at analyzing the gap between the achievable rate and the information-theoretic lower bounds on the optimal rate.
To this end, we proceed in two main steps.
First, we give some preliminary results in Appendix~\ref{app:gap-proof-preliminaries} that will be useful for the main body of the proof.
These include some in-depth analysis of the achievable rate, as well as some generally-useful definitions and inequalities.
Second, we present the gap analysis itself in Appendix~\ref{app:gap-proof-analysis}.
We identify the different cases to analyze, and prove the order-optimality of the scheme in each case.

For clarity, the proofs of the preliminary results are relegated to the end of this appendix, in Appendix~\ref{app:gap-proof-prelims-proof}.
Finally, we separately analyze, in Appendix~\ref{app:gap-proof-smallK}, a special case where the total number of caches is smaller than some constant.

\subsection{Preliminary results}\label{app:gap-proof-preliminaries}

We will now present the preliminary results that will be used for the gap analysis. The proofs of all the lemmas in this section are relegated to Appendix~\ref{app:gap-proof-prelims-proof}.

For convenience, we start by defining, for any set of levels $A\subseteq\{1,\ldots,L\}$:
\begin{IEEEeqnarray*}{rCl}
S_A &=& \sum_{i\in A} \sqrt{N_iU_i},\\
T_A &=& \sum_{i\in A} \frac{N_i}{d_i}.
\end{IEEEeqnarray*}

Recall that the achievability scheme partitions the set of levels into an $M$-feasible partition (Definition~\ref{def:m-feasible}).
This provides us with a structured way of studying the behavior of the individual rates of different levels.
However, in some cases, a particular level in the set $I$ might have so much allocated memory that its rate, while still positive, becomes small enough to obfuscate the properties of its other parameters, such as number of files and number of users.
This prompts us to isolate such levels and study their behavior separately from the remaining levels in $I$.
To this end, we refine the $M$-feasible partition currently in use by defining:
\[
I'' = \left\{ i\in I : \alpha_iM > \beta\frac{N_i}{d_i} \right\} = \left\{ i\in I : M > \frac{\beta}{d_i}\sqrt{\frac{N_i}{U_i}}S_I+T_J \right\},
\]
where $\beta<1$ is a constant defined by:
\begin{equation}\label{eq:beta}
\beta = \frac{1}{16\gamma(D+1)L},
\end{equation}
and $\gamma = \frac{1}{1-e^{-1}}$. The complement of $I''$, with respect to $I$, is denoted by $I' = I\backslash I''$.

As a result (recall Definition~\ref{def:m-feasible}), the memory satisfies the following inequalities at any time:
\begin{IEEEeqnarray}{lCl"rCcCl}
\IEEEyesnumber\label{eq:m-conditions}
\IEEEyessubnumber*
\forall h&\in& H,& && M &<& \frac{1}{K}\sqrt{\frac{N_h}{U_h}}S_I+T_J + \frac{N_h}{K}, \label{eq:m-conditions-h}\\
\forall i&\in& I',&  \frac{1}{K}\sqrt{\frac{N_i}{U_i}}S_I+T_J &\le& M &<& \frac{\beta}{d_i}\sqrt{\frac{N_i}{U_i}}S_I + T_J, \label{eq:m-conditions-i}\\
\forall i&\in& I'',& \frac{\beta}{d_i}\sqrt{\frac{N_i}{U_i}}S_I+T_J &\le& M &<& \frac{1}{d_i}\sqrt{\frac{N_i}{U_i}}S_I+T_J, \label{eq:m-conditions-i1}\\
\forall j&\in& J,&   \frac{1}{d_i}\sqrt{\frac{N_j}{U_j}}S_I+T_J &\le& M. \label{eq:m-conditions-j}
\end{IEEEeqnarray}

From the above definitions, as well as from the PAMA choice of memory-sharing parameters $\{\alpha_i\}$, we can deduce the individual broadcast rate of each level.
This is done in the next lemma.
\begin{lemma}\label{lemma:achievability}
The individual rates of each level can be bounded as in the following inequalities:
\begin{IEEEeqnarray*}{lCl"rCcCl}
\forall h&\in& H,& && R_h(M) &=& KU_h,\\
\forall i&\in& I',&  (1-e^{-1})\frac{S_I\sqrt{N_iU_i}}{M-T_J} &\le& R_i(M) &\le& \frac{S_I\sqrt{N_iU_i}}{M-T_J}-d_iU_i,\\
\forall i&\in& I'',& (1-e^{-1}) d_iU_i\left(1-\frac{M-T_J}{\frac{1}{d_i}\sqrt{\frac{N_i}{U_i}}S_I}\right) &\le& R_i(M) &\le& \frac1\beta d_iU_i\left(1-\frac{M-T_J}{\frac{1}{d_i}\sqrt{\frac{N_i}{U_i}}S_I}\right),\\
\forall j&\in& J,&   && R_j(M) &=& 0.
\end{IEEEeqnarray*}
\end{lemma}

The next step is to identify two important levels, each of which will dominate one of two significant sums.
The first one, labeled $m$, is the level that dominates the rate achieved by the scheme.
In particular, it is the level for which the individual rate is the largest.
Recall that the individual rate of level $i$ is $R_i(M) = R^\text{SL}(\alpha_iM,K,N_i,U_i,d_i)$.
The second one, labeled $i^\ast$, is the level that dominates the sum $S_{I'}=\sum_{i\in I'}\sqrt{N_iU_i}$.
These definitions are formalized next.
\begin{IEEEeqnarray*}{rCl}
m &=& \argmax_i R_i(M),\\
i^\ast &=& \argmax_{i\in I'} N_iU_i. \IEEEyesnumber\label{eq:special-levels}
\end{IEEEeqnarray*}

Throughout the analysis, the total broadcast rate will be upper-bounded by the individual rate of level $m$:
\begin{equation}\label{eq:rate-m}
R(M) = \sum_{i=1}^L R_i(M) \le L \cdot R_m(M).
\end{equation}

In addition, the properties of the achievable rate are such that level $m$ also often dominates the $\{N_iU_i\}_i$ terms, as the following lemma states.
\begin{lemma}[Dominance Lemma]\label{lemma:dominance}
If $m\in H\cup I'$, then $S_{I'} \le \gamma L \sqrt{N_mU_m}$, with $\gamma=\frac{1}{1-e^{-1}}$.
\end{lemma}

Next, we revisit the regularity condition in \eqref{eq:regularity-level-separation}. We rewrite it in a format that will be more useful later on:
\begin{equation}\label{eq:regularity-level-separation-2}
\forall i<j,\quad \frac{\sqrt{U_i/N_i}}{\sqrt{U_j/N_j}} \ge q_0 = 16\gamma(D+1)^2L,
\end{equation}
where $\gamma=\frac{1}{1-e^{-1}}$ as above. Note that $q_0=\sqrt{q}$ in \eqref{eq:regularity-level-separation}.
Notice that the regularity condition implies that:
\begin{equation}\label{eq:order-preserved}
\forall i,j,\quad
\sqrt{\frac{N_i}{U_i}} < \sqrt{\frac{N_j}{U_j}}
\implies
\frac{1}{d_i}\sqrt{\frac{N_i}{U_i}} < \frac{1}{d_j}\sqrt{\frac{N_j}{U_j}}.
\end{equation}
This is effectively saying that, for any $M$-feasible partition $(H,I,J)$ with refining $I=I'\cup I''$, the levels in $J$ will always be more popular than the levels in $I''$, which are in turn more popular than the ones in $I'$, and $H$ contains the least popular levels of all.
This can be written concisely as $J<I''<I'<H$.

The following lemma is a direct consequence of the regularity condition.
\begin{lemma}\label{lemma:level-spacing}
For all values of $M\ge0$, the ML-PAMA scheme always results in $|I''|\le1$.
\end{lemma}

Finally, we present two bounds that will frequently be used in the gap analysis: an upper bound on the achievable rate; and a lower bound on the optimal rate.

\begin{lemma}\label{lemma:uba}
If $I''=\{i_1\}\not=\emptyset$, then, for all $i\in H\cup I'$,
\[
R_i(M) \le \frac{d_{i_1}}{\beta} \sqrt{\frac{N_iU_iU_{i_1}}{N_{i_1}}}.
\]
\end{lemma}

\begin{lemma}\label{lemma:lbb}
If $I''=\{i_1\}\not=\emptyset$, and if $I'\not=\emptyset$ so that $i^\ast$ (as defined in \eqref{eq:special-levels}) exists, then,
\[
R^\ast(M) \ge \frac{1}{16L(D+1)^2} \sqrt{\frac{N_{i^\ast}U_{i^\ast}U_{i_1}}{N_{i_1}}},
\]
provided $q_0\ge16L(D+1)^2$.
\end{lemma}

\subsection{Gap analysis}\label{app:gap-proof-analysis}

The gap analysis consists in comparing the achievable rate with the lower bounds given in Lemma~\ref{lemma:converse-cutset} and Lemma~\ref{lemma:converse-general}.
The lower bounds in Lemma~\ref{lemma:converse-general} require the evaluation of two minimizations. For clarity, we label the two minimizations:
\begin{IEEEeqnarray}{rCl}
\IEEEyesnumber\label{eq:minimize}
\IEEEyessubnumber*
\label{eq:minimize-l}
s(s-d_l+1)b &\lessgtr& \frac{N_l}{U_l};\\
\label{eq:minimize-A}
b &\lessgtr& \frac{N_j}{d_jU_j},\quad \forall j\in A.
\end{IEEEeqnarray}

Furthermore, we will use, on one occasion, the following result, which is a direct consequence of Lemma~\ref{lemma:converse-general}.
\begin{corollary}[Simple general bounds]\label{coroll:converse-general}
For any $A\subseteq\{1,\ldots,L\}$ and any $b\in\mathbb{N}^+$, we have:
\[
R^\ast(M) \ge \sum_{j\in A} \min\left\{ U_j \,,\, \frac{N_j}{bd_j} \right\} - \frac{M}{b}.
\]
\end{corollary}
Corollary~\ref{coroll:converse-general} requires only the evaluation of the minimization in \eqref{eq:minimize-A}.

The analysis in this section is concerned with the case when the number of caches $K$ in the network is large.
In particular, we assume:
\begin{equation}\label{eq:k}
K \ge k_0 = 16(D+1)^2(\gamma L+1),
\end{equation}
where $\gamma=\frac{1}{1-e^{-1}}$. The case where $K<k_0$ will be dealt with in Appendix~\ref{app:gap-proof-smallK}.

We will divide the analysis of this section into two main regimes, which depend on the set to which $m$ belongs.
Note that, if $m\in J$, then $R_m(M)=0$ and thus $R(M)=0$.
Thus, we will only consider the regimes $m\in H\cup I'$ and $m\in I''$.
The first regime is subdivided into two cases, depending on whether $I''$ is empty or not.
The second regime is also subdivided into two cases, depending on whether $I'$ is empty or not.

\subsubsection{Regime 1: $m\in H\cup I'$}

\paragraph{Case A: $I''=\{i_1\}$}~\\
From Lemma~\ref{lemma:uba} and \eqref{eq:rate-m}, we have:
\begin{equation}\label{eq:ach1}
R(M) \le L\cdot R_m(M) \le L\cdot \frac{D+1}{\beta}\sqrt{\frac{N_mU_mU_{i_1}}{N_{i_1}}}.
\end{equation}
Combining \eqref{eq:ach1} with Lemma~\ref{lemma:lbb}, we get:
\begin{equation}\label{eq:gap1}
\frac{R(M)}{R^\ast(M)} \le \frac{16}{\beta}L^2(D+1)^3.
\end{equation}

\paragraph{Case B: $I''=\emptyset$}~\\
Notice that $I'=I$ and hence $S_I=S_{I'}$.
Since $m\in H\cup I'$, we have, from Lemma~\ref{lemma:dominance},
\begin{equation}\label{eq:dominance-full}
S_I \le \gamma L \sqrt{N_mU_m}.
\end{equation}

\textbf{Case B.i: $m\in H$}~\\
Here, we have
\begin{equation}\label{eq:ach2}
R(M) \le L\cdot KU_m.
\end{equation}

\emph{Consider the case where $J\not=\emptyset$.} Use Lemma~\ref{lemma:converse-general} with $l=m$, $A=J$, $s=\floor{\delta K}$, $b=\floor{\frac{1}{\delta^2K^2}\cdot\frac{N_m}{U_m}}$, and
\[
\delta = \frac{1}{4(D+1)(\gamma L+1)}.
\]

We must first analyze the inequalities in \eqref{eq:minimize-l} and \eqref{eq:minimize-A}.
For \eqref{eq:minimize-l}, we have:
\[
s(s-d_l+1)b \le s^2b \le \delta^2K^2 \cdot \frac{1}{\delta^2K^2}\cdot\frac{N_m}{U_m} = \frac{N_m}{U_m}.
\]
For \eqref{eq:minimize-A}, we first show that the quantity inside the floor in the expression of $b$ is greater than 1.
Indeed, since $J$ is non-empty, there exists some $j\in J$.
Therefore, since $m\in H$ and $j\in J$, we have:
\[
\frac{1}{d_j}\sqrt{\frac{N_j}{U_j}} \le \tilde M \le \frac{1}{K}\sqrt{\frac{N_m}{U_m}} \implies \frac{N_m}{K^2U_m} \ge \frac{N_j}{d_j^2U_j},
\]
Hence,
\begin{equation}
\label{eq:b-1}
\frac{1}{\delta^2K^2}\cdot\frac{N_m}{U_m}
= (D+1)64\gamma ^2L^2\cdot\frac{N_m}{K^2U_m}
\ge 2\cdot\frac{(D+1)N_m}{K^2U_m} \ge 2\cdot\frac{(D+1)N_j}{d_j^2U_j} \ge 2\cdot\frac{N_j}{d_jU_j} \ge 1.
\end{equation}
As a result of \eqref{eq:b-1}, $b\ge\frac12\cdot\frac{1}{\delta^2K^2}\cdot\frac{N_m}{U_m} \ge \frac12\cdot2\cdot\frac{N_j}{d_jU_j}=\frac{N_j}{d_jU_j}$ for all $j\in J$, thus evaluating the inequality in \eqref{eq:minimize-A}.

Therefore,
\begin{IEEEeqnarray*}{rCl}
R^\ast(M)
&\ge& \frac{1}{D+1} (s-d_m+1)U_m - \frac{M-T_J}{b}\\
&\ge& \frac{1}{D+1} \left( \delta K-1 - d_m+1\right)U_m - \frac{M-T_J}{\frac{1}{2\delta^2K^2}\cdot\frac{N_m}{U_m}}\\
&=& \frac{1}{D+1}\left(\delta K-d_m\right)U_m - 2\delta^2\cdot \frac{M-T_J}{\frac{N_m}{K^2U_m}}\\
&\overset{(a)}{\ge}& \frac{1}{D+1}\left(\delta K-d_m\right)U_m - 2\delta^2\cdot \frac{\frac{1}{K}\sqrt{\frac{N_m}{U_m}}\cdot S_I+\frac{N_m}{K}}{\frac{N_m}{K^2U_m}}\\
&\overset{(b)}{\ge}& \frac{1}{D+1}\left(\delta K-d_m\right)U_m - 2\delta^2\cdot \frac{\frac{1}{K}\sqrt{\frac{N_m}{U_m}}\cdot \gamma L\sqrt{N_mU_m}+\frac{N_m}{K}}{\frac{N_m}{K^2U_m}}\\
&=& \frac{1}{D+1}\left(\delta K-d_m\right)U_m - 2\delta^2(\gamma L+1) \cdot KU_m\\
&=& KU_m \left[ \frac{\delta}{D+1} - \frac{d_m}{K(D+1)} - 2\delta^2(\gamma L+1) \right]\\
&=& KU_m \left[ \frac{1}{(D+1)^2 4(\gamma L+1)} - \frac{d_m}{K(D+1)} - \frac{2(\gamma L+1)}{(D+1)^2 16(\gamma L+1)^2} \right]\\
&=& KU_m\left[ \frac{1}{(D+1)^2 8(\gamma L+1)} - \frac{d_m}{K(D+1)} \right]\\
&\ge& KU_m\left[ \frac{1}{(D+1)^2 8(\gamma L+1)} - \frac1K \right]\\
&\overset{(c)}{\ge}& KU_m\cdot\frac{1}{(D+1)^2 16(\gamma L+1)}, \IEEEyesnumber\label{eq:conv2a}
\end{IEEEeqnarray*}
where $(a)$ follows from $m\in H$ and \eqref{eq:m-conditions-h}, $(b)$ follows from \eqref{eq:dominance-full}, and $(c)$ follows from \eqref{eq:k}.

We can now combine \eqref{eq:ach2} with \eqref{eq:conv2a} to get:
\begin{equation}\label{eq:gap2a}
\frac{R(M)}{R^\ast(M)} \le \frac{L}{\frac{1}{(D+1)^2 16(\gamma L+1)} } = 16(D+1)^2L(\gamma L+1).
\end{equation}

\emph{Now consider the case when $J=\emptyset$.} Hence, $T_J=T_\emptyset=0$.
We will use Lemma~\ref{lemma:converse-cutset}, with $i=m$, and $v=\floor{\delta KU_m}$, with:
\[
%\delta = 1 - \sqrt{ \frac{2\gamma L}{2\gamma L+1} }.
\delta = \frac{1}{4(\gamma L+1)}.
\]
Notice that, from \eqref{eq:k}, we have $K\ge4(\gamma L+1)D = \frac{D}{\delta} \ge \frac{d_m}{\delta}$, which ensures that $v\ge d_m$.
Then,
\begin{IEEEeqnarray*}{rCl}
R^\ast(M)
&\ge& v - \frac{ \ceil{v/U_m} + (d_m-1) }{ \floor{N_m/v} }M\\
&\ge& \delta KU_m - 1 - \frac{ \delta K + d_m }{ \frac{N_m}{\delta KU_m} -1 }M\\
&=& \delta KU_m - 1 - \frac{ \delta KU_m\left( \delta K + d_m \right) }{ N_m - \delta KU_m }M\\
&\overset{(a)}{\ge}& \delta KU_m - 1 - \frac{\delta KU_m\left(\delta K+d_m\right)}{(1-\delta)N_m} \cdot \left( \frac{1}{K}\sqrt{\frac{N_m}{U_m}}S_I + \frac{N_m}{K} \right)\\
&=& \delta KU_m - 1 - \frac{\delta KU_m\left(\delta + \frac{d_m}{K}\right)}{(1-\delta)N_m} \cdot \left( \sqrt{\frac{N_m}{U_m}}\cdot S_I + N_m \right)\\
&\overset{(b)}{\ge}& \delta KU_m - 1 - \frac{\delta KU_m\left(\delta + \frac{d_m}{K}\right)}{(1-\delta)N_m} \cdot \left( \sqrt{\frac{N_m}{U_m}}\cdot\gamma L\sqrt{N_mU_m} + N_m \right)\\
&=& \delta KU_m - 1 - \frac{ (\gamma L+1) \cdot \delta KU_m \left(\delta+\frac{d_m}{K}\right) }{ 1-\delta }\\
&=& KU_m\left[ \delta - \frac{1}{KU_m} - \frac{(\gamma L+1)\delta\left(\delta+\frac{d_m}{K}\right)}{1-\delta} \right]\\
&\overset{(c)}{\ge}& KU_m\left[ \delta - \frac1K - \frac{2(\gamma L+1)\delta^2}{1-\delta} \right]\\
&=& KU_m\left[ \frac{1}{4(\gamma L+1)} - \frac1K - \frac{2(\gamma L+1) \cdot\frac{1}{16(\gamma L+1)^2}}{1-\frac{1}{4(\gamma L+1)}} \right]\\
&=& KU_m\left[ \frac{1}{4(\gamma L+1)} - \frac1K - \frac{2}{16(\gamma L+1) - 4} \right]\\
&\ge& KU_m\left[ \frac{1}{4(\gamma L+1)} - \frac1K - \frac{2}{12(\gamma L+1)} \right]\\
&=& KU_m\left[ \frac{1}{12(\gamma L+1)} - \frac1K \right]\\
&\overset{(d)}{\ge}& KU_m \cdot \frac{1}{24(\gamma L+1)}, \IEEEyesnumber \label{eq:conv2b}
\end{IEEEeqnarray*}
where $(a)$ uses $m\in H$ and \eqref{eq:m-conditions-h}, $(b)$ uses \eqref{eq:dominance-full}, $(c)$ uses $K\ge d_m/\delta$ and $U_m\ge1$, and $(d)$ uses $K\ge k_0\ge 24(\gamma L+1)$.

We then combine \eqref{eq:ach2} with \eqref{eq:conv2b} to get:
\begin{equation}\label{eq:gap2b}
\frac{R(M)}{R^\ast(M)} \le 24L(\gamma L+1).
\end{equation}

\textbf{Case B.ii: $m\in I'$.}~\\
Using \eqref{eq:rate-m}, Lemma~\ref{lemma:achievability} and \eqref{eq:dominance-full}, the achievable rate can be bounded by:
\begin{equation}\label{eq:ach3}
R(M) \le L\cdot R_m(M) \le L \cdot \frac{S_I\sqrt{N_mU_m}}{M-T_J} \le \frac{\gamma L^2N_mU_m}{M-T_J}.
\end{equation}

\emph{Look at the case when $J\not=\emptyset$.} Use Lemma~\ref{lemma:converse-general}, with $l=m$, $A=J$, $s=\floor{\frac{\delta N_m}{M-T_J}}$, $b=\floor{\frac{(M-T_J)^2}{\delta^2N_mU_m}}$, and:
\[
\delta = \frac{1}{4(D+1)}.
\]
We will now analyze the minimizations in \eqref{eq:minimize-l} and \eqref{eq:minimize-A}.
For \eqref{eq:minimize-l}:
\[
s(s-d_m+1)b \le s^2b \le \frac{\delta^2N_m^2}{(M-T_J)^2} \cdot \frac{(M-T_J)^2}{\delta^2N_mU_m} = \frac{N_m}{U_m}.
\]
For \eqref{eq:minimize-A}, note that there exists some $j\in J$.
Hence, using \eqref{eq:m-conditions-j}:
\begin{IEEEeqnarray*}{rCl}
\frac{(M-T_J)^2}{\delta^2N_mU_m}
&\ge& \frac{1}{\delta^2N_mU_m} \left( \frac{1}{d_j}\sqrt{\frac{N_j}{U_j}}S_I \right)^2 = \frac{S_I^2}{\delta^2N_mU_m} \cdot \frac{1}{d_j^2}\cdot\frac{N_j}{U_j} \ge \frac{1}{\delta^2d_j}\cdot\frac{N_j}{d_jU_j}\\
&=& \frac{16(D+1)^2}{d_j}\cdot\frac{N_j}{d_jU_j} \ge 2\cdot\frac{N_j}{d_jU_j}.
\end{IEEEeqnarray*}
Furthermore, since $\frac{N_j}{d_jU_j}\ge1$, then,
\[
b \ge \frac12 \cdot \frac{(M-T_J)^2}{\delta^2N_mU_m} \ge \frac{N_j}{d_jU_j}.
\]
Therefore,
\begin{IEEEeqnarray*}{rCl}
R^\ast(M)
&\ge& \frac{1}{D+1}(s-(d_m-1))U_m - \frac{M-T_J}{b}\\
&\ge& \frac{1}{D+1}\left(\frac{\delta N_mU_m}{M-T_J} - d_mU_m\right) - \frac{M-T_J}{\frac{(M-T_J)^2}{2\delta^2N_mU_m}}\\
&=& \frac{1}{D+1}\cdot\frac{\delta N_mU_m}{M-T_J} - \frac{d_mU_m}{D+1} - \frac{2\delta^2N_mU_m}{M-T_J}\\
&=& \frac{N_mU_m}{M-T_J} \left[ \frac{\delta}{D+1} - \frac{M-T_J}{\frac{N_m}{d_m}(D+1)} - 2\delta^2 \right]\\
&=& \frac{N_mU_m}{M-T_J} \left[ \frac{1}{4(D+1)^2} - \frac{M-T_J}{\frac{N_m}{d_m}(D+1)} - \frac{2}{16(D+1)^2} \right]\\
&\overset{(a)}{\ge}& \frac{N_mU_m}{M-T_J} \left[ \frac{1}{8(D+1)^2} - \frac{1}{\frac{N_m}{d_m}(D+1)}\cdot\frac{\beta}{d_m}\sqrt{\frac{N_m}{U_m}}S_I \right]\\
&\overset{(b)}{\ge}& \frac{N_mU_m}{M-T_J} \left[ \frac{1}{8(D+1)^2} - \frac{\beta\gamma L}{D+1} \right]\\
&\overset{(c)}{=}& \frac{N_mU_m}{M-T_J} \left[ \frac{1}{8(D+1)^2} - \frac{1}{16(D+1)^2} \right]\\
&=& \frac{N_mU_m}{M-T_J} \cdot \frac{1}{16(D+1)^2} , \IEEEyesnumber\label{eq:conv3a}
\end{IEEEeqnarray*}
where $(a)$ follows from $m\in I'$ and \eqref{eq:m-conditions-i}, $(b)$ follows from \eqref{eq:dominance-full}, and $(c)$ follows from \eqref{eq:beta}.

We now combine \eqref{eq:ach3} with \eqref{eq:conv3a} to get:
\begin{equation}\label{eq:gap3a}
\frac{R(M)}{R^\ast(M)} \le \frac{(D+1)\gamma L^2}{\frac{1}{8(D+1)}-\beta\gamma L} \le 16(D+1)^2\gamma L^2.
\end{equation}

\emph{Now assume $J=\emptyset$.} Note that $T_J=0$.
Use Lemma~\ref{lemma:converse-cutset}, using $i=m$ and $v=\floor{\frac{\delta N_mU_m}{M}}$, with
\[
\delta = 1 - \sqrt{\frac23}.
\]
We have,
\begin{IEEEeqnarray*}{rCl}
R^\ast(M)
&\ge& \frac{\delta N_mU_m}{M} - 1 - \frac{ \frac{\delta N_m}{M} + d_m }{ \frac{M}{\delta U_m} - 1 }M\\
&\ge& \frac{\delta N_mU_m}{M} - 1 - \delta U_m\cdot\frac{ \frac{\delta N_m}{M} + d_m }{ M - \delta U_m }M. \IEEEyesnumber\label{eq:conv3b-step1}
\end{IEEEeqnarray*}
Note the following:
\[
\frac{\delta N_m}{M} \ge \frac{\delta N_m}{\frac{\beta}{d_m}\sqrt{\frac{N_m}{U_m}}S_I} = \frac{\delta d_m \sqrt{N_mU_m}}{\beta S_I} \ge d_m\cdot \frac{\delta}{\beta\gamma L} \ge d_m,
\]
where the last inequality holds because $\beta < \frac{\delta}{\gamma L}$ follows from \eqref{eq:beta}.
Also note that, from \eqref{eq:m-conditions-i}:
\[
M \ge \frac{1}{K}\sqrt{\frac{N_m}{U_m}}S_I \ge \frac{1}{K} N_m \ge U_m.
\]

Substituting the above two inequalities in \eqref{eq:conv3b-step1}:
\begin{IEEEeqnarray*}{rCl}
R^\ast(M)
&\ge& \frac{\delta N_mU_m}{M} - 1 - \delta U_m \cdot \frac{ \frac{2\delta N_m}{M} }{ M-\delta M } M\\
&=& \frac{\delta N_mU_m}{M} - 1 - \frac{N_mU_m}{M}\cdot\frac{2\delta^2}{1-\delta}\\
&=& \frac{N_mU_m}{M} \left[ \delta - \frac{M}{N_mU_m} - \frac{2\delta^2}{1-\delta} \right]\\
&=& \frac{N_mU_m}{M} \left[ 5-2\sqrt{6} - \frac{M}{N_mU_m} \right]\\
&\overset{(a)}{\ge}& \frac{N_mU_m}{M} \left[ 0.1 - \frac{\frac{\beta}{d_m}\sqrt{\frac{N_m}{U_m}}S_I}{N_mU_m} \right]\\
&\overset{(b)}{\ge}& \frac{N_mU_m}{M} \left[ 0.1 - \frac{\frac{\beta}{d_m}\gamma L}{U_m} \right]\\
&\overset{(c)}{\ge}& \frac{N_mU_m}{M} \left[ 0.1 - \beta\gamma L \right]\\
&\overset{(d)}{\ge}& 0.05\frac{N_mU_m}{M}, \IEEEyesnumber\label{eq:conv3b}
\end{IEEEeqnarray*}
where $(a)$ follows from $m\in I'$ and \eqref{eq:m-conditions-i}, $(b)$ follows from \eqref{eq:dominance-full}, $(c)$ follows from $d_mU_m\ge1$, and $(d)$ follows from $\beta<\frac{0.05}{\gamma L}$, which is implied by \eqref{eq:beta}.

By combining \eqref{eq:ach3} (with $T_J=0$) with \eqref{eq:conv3b}, we get:
\begin{equation}\label{eq:gap3b}
\frac{R(M)}{R^\ast(M)} \le 20\gamma L^2.
\end{equation}

\subsubsection{Regime 2: $m\in I''$}

\paragraph{Case A: $I'=\emptyset$}~\\
When $I'=\emptyset$, then $I=\{m\}$, and $S_I=\sqrt{N_mU_m}$.
Therefore, we have:
\begin{equation*}
R_m(M) \le \frac1\beta d_mU_m\left(1-\frac{M-T_J}{\frac{1}{d_m}\sqrt{\frac{N_m}{U_m}}S_I}\right)
= \frac1\beta d_mU_m\left( 1 - \frac{M-T_J}{N_m/d_m} \right),
\end{equation*}
and hence
\begin{equation}\label{eq:ach-i0}
R(M) \le L\cdot\frac1\beta (D+1)U_m \left( 1 - \frac{M-T_J}{N_m/d_m}\right).
\end{equation}

We will now use Corollary~\ref{coroll:converse-general}, with $A=J\cup\{m\}$ and $b=\ceil{\frac{N_m}{d_mU_m}}$.
Evaluating \eqref{eq:minimize-A}:
\[
b \ge \frac{N_m}{d_mU_m} \ge \frac{q_0^2N_j}{d_mU_j} \ge \frac{N_j}{U_j} \ge \frac{N_j}{d_jU_j},
\]
for all $j\in J$.
Note also that $b\ge1$, and thus $b\le\frac{2N_m}{d_mU_m}$.
Therefore:
\begin{IEEEeqnarray*}{rCl}
R^\ast(M)
&\ge& \frac{M-T_{J\cup\{m\}}}{b}\\
&\ge& \frac{ \frac{N_m}{d_m} - \left(M-T_J\right) }{ \frac{2N_m}{d_mU_m} }\\
&=& \frac{U_m}{2} \left( 1 - \frac{M-T_J}{N_m/d_m} \right). \IEEEyesnumber\label{eq:conv-linear}
\end{IEEEeqnarray*}

Combining \eqref{eq:conv-linear} with \eqref{eq:ach-i0}, as well as using \eqref{eq:beta}, we get:
\begin{equation}
\frac{R(M)}{R^\ast(M)} \le \frac{2(D+1)L}{\beta} = 32\gamma(D+1)^2L^2.
\end{equation}

\paragraph{Case B: $I'\not=\emptyset$}~\\
Note that when $I'\not=\emptyset$, the level $i^\ast$ exists.

By the level-spacing regularity condition, $|I''|\le1$, which implies $I''=\{m\}$ in this regime.% We will subdivide this regime into three cases.

Consider the following two memory values:
\begin{align*}
Y_0 &= \frac{N_m}{d_m} + T_J,\\
Y_1 &= \frac{\beta}{d_m}\sqrt{\frac{N_m}{U_m}}S_I + T_J.
\end{align*}
Note that $M\ge Y_1$ because $m\in I''$ and \eqref{eq:m-conditions-i1}.

We will analyze two cases, that depend on the value of $Y_0-Y_1$.

\textbf{Case B.i: $Y_0-Y_1 < \frac{N_m}{2d_m}$}~\\
Note the following:
\begin{IEEEeqnarray*}{rCl}
Y_0 - Y_1 &<& \frac{N_m}{2d_m}\\
\frac{N_m}{d_m} - \frac{\beta}{d_m}\sqrt{\frac{N_m}{U_m}}S_I &<& \frac{N_m}{2d_m}\\
\frac{\beta}{d_m}\sqrt{\frac{N_m}{U_m}}S_I &>& \frac{N_m}{2d_m}\\
2\beta \cdot \frac{S_I}{\sqrt{N_mU_m}} &>& 1.
\end{IEEEeqnarray*}
By noticing that, because $m\in I''$, then $S_I=\sqrt{N_mU_m}+S_{I'} \le \sqrt{N_mU_m}+L\sqrt{N_{i^\ast}U_{i^\ast}}$ (by the definition of $i^\ast$, see \eqref{eq:special-levels}), the above implies:
\begin{IEEEeqnarray*}{rCl}
2\beta\left( 1 + L\sqrt{\frac{N_{i^\ast}U_{i^\ast}}{N_mU_m}} \right) &>& 1\\
L\sqrt{\frac{N_{i^\ast}U_{i^\ast}}{N_mU_m}} &>& \frac{1}{2\beta}-1\\
\frac{2\beta}{1-2\beta}L\sqrt{\frac{N_{i^\ast}U_{i^\ast}U_m}{N_m}} &>& U_m. \IEEEyesnumber\label{eq:ach-step}
\end{IEEEeqnarray*}

Since $M\ge Y_1$, we have, by Lemma~\ref{lemma:achievability} and \eqref{eq:ach-step}:
\[
R_m(M) \le R_m(Y_1) \le \frac1\beta d_mU_m \le \frac{2}{1-2\beta}\cdot d_m \cdot L\sqrt{\frac{N_{i^\ast}U_{i^\ast}U_m}{N_m}},
\]
and thus
\begin{equation}\label{eq:ach-y1}
R(M) \le \frac{2(D+1)L^2}{1-2\beta} \sqrt{\frac{N_{i^\ast}U_{i^\ast}U_m}{N_m}}.
\end{equation}

Combining \eqref{eq:ach-y1} with Lemma~\ref{lemma:lbb} (where $i_1=m$), we get:
\begin{equation}
\frac{R(M)}{R^\ast(M)} \le \frac{32(D+1)^3L^3}{1-2\beta} \le \frac87\cdot32(D+1)^3L^3 \le 37(D+1)^3L^3,
\end{equation}
which follows from observing that:
\[
1-2\beta = 1-\frac{1}{8\gamma(D+1)L} = \frac{8\gamma(D+1)L-1}{8\gamma(D+1)L} \ge \frac78.
\]

\textbf{Case B.ii: $Y_0-Y_1 \ge \frac{N_m}{2d_m}$}~\\
We consider two possibilities: one where $M$ is greater than $Y_0$, and one where it is less than $Y_0$.

When $M\ge Y_0$, we have:
\begin{equation}\label{eq:ach-gty0}
R_m(M) \le R_m(Y_0).
\end{equation}

When $M<Y_0$, we have $Y_1\le M<Y_0$, and hence, by the convexity of $R_m(\cdot)$:
\begin{IEEEeqnarray*}{rCl}
R_m(M) &\le& R_m(Y_0) + \frac{Y_0-M}{Y_0-Y_1}\left(R_m(Y_1)-R_m(Y_0)\right)\\
&\le& R_m(Y_0) + \frac{Y_0-M}{Y_0-Y_1}R_m(Y_1)\\
&\le& R_m(Y_0) + \frac{Y_0-M}{\frac{N_m}{2d_m}} \cdot \frac1\beta d_mU_m\\
&=& R_m(Y_0) + \frac2\beta d_mU_m \left( 1 - \frac{M-T_J}{N_m/d_m} \right). \IEEEyesnumber\label{eq:ach-lty0}
\end{IEEEeqnarray*}

We now need to compute $R_m(Y_0)$ for both \eqref{eq:ach-gty0} and \eqref{eq:ach-lty0}.
From Lemma~\ref{lemma:achievability}:
\begin{IEEEeqnarray*}{rCl}
R_m(Y_0)
&\le& \frac{S_I\sqrt{N_mU_m}}{Y_0-T_J} - d_mU_m\\
&=& \frac{S_I\sqrt{N_mU_m}}{N_m/d_m} - d_mU_m\\
&=& d_m\sqrt{\frac{U_m}{N_m}}S_I - d_mU_m\\
&=& d_m\sqrt{\frac{U_m}{N_m}}\left(\sqrt{N_mU_m}+S_{I'}\right) - d_mU_m\\
&=& d_m\sqrt{\frac{U_m}{N_m}}S_{I'}\\
&\le& d_m L \sqrt{\frac{N_{i^\ast}U_{i^\ast}U_m}{N_m}}. \IEEEyesnumber\label{eq:ach-y0}
\end{IEEEeqnarray*}

Therefore, when $M\ge Y_0$, we have:
\[
R(M) \le L\cdot R_m(M) \le L\cdot R_m(Y_0) \le (D+1)L^2\sqrt{\frac{N_{i^\ast}U_{i^\ast}U_m}{N_m}}.
\]
This can be combined with Lemma~\ref{lemma:lbb} to give:
\begin{equation}\label{eq:gap4a}
\frac{R(M)}{R^\ast(M)} \le 16L^3(D+1)^3.
\end{equation}

When $M<Y_0$, then:
\begin{equation}
R(M) \le (D+1)L^2\sqrt{\frac{N_{i^\ast}U_{i^\ast}U_m}{N_m}} + \frac2\beta \cdot(D+1)L\cdot U_m\left(1-\frac{M-T_J}{N_m/d_m} \right).
\end{equation}
Define:
\begin{align*}
r_1 &= \sqrt{\frac{N_{i^\ast}U_{i^\ast}U_m}{N_m}}\\
r_2 &= U_m\left(1-\frac{M-T_J}{N_m/d_m}\right),
\end{align*}
and thus we can rewrite the rate as:
\[
R(M) \le (D+1)L^2\cdot r_1 + \frac{2(D+1)L}{\beta}\cdot r_2.
\]
Also define:
\[
\lambda = 8(D+1)^2L
\]

If $r_1\le\lambda r_2$, then:
\begin{IEEEeqnarray*}{rCl}
R(M) &\le& \left[ (D+1)L^2\lambda + \frac{2(D+1)L}{\beta} \right]r_2\\
&=& \left[ 8(D+1)^3L^3 + \frac{2(D+1)L}{\beta} \right] U_m\left(1-\frac{M-T_J}{N_m/d_m}\right).
\end{IEEEeqnarray*}
Using the lower bound computed above in \eqref{eq:conv-linear}, we get:
\begin{equation}\label{eq:gap4b1}
\frac{R(M)}{R^\ast(M)} \le 2(D+1)L\left[ 8(D+1)^2L^2 + \frac2\beta \right].
\end{equation}

If $r_1>\lambda r_2$, then:
\begin{IEEEeqnarray*}{rCl}
R(M) &\le& \left[ (D+1)L^2 + \frac{2(D+1)L}{\beta} \cdot \frac{1}{\lambda} \right] r_1\\
&=& \left[ (D+1)L^2 + \frac{2}{8\beta(D+1)} \right] \sqrt{\frac{N_{i^\ast}U_{i^\ast}U_m}{N_m}}.
\end{IEEEeqnarray*}
Combining this with Lemma~\ref{lemma:lbb}, we get:
\begin{equation}\label{eq:gap4b2}
\frac{R(M)}{R^\ast(M)} \le 2(D+1)L\left[ 8(D+1)^2L^2 + \frac2\beta \right].
\end{equation}

Therefore, both \eqref{eq:gap4b1} and \eqref{eq:gap4b2} give, using \eqref{eq:beta} and the fact that $\gamma<2$:
\begin{equation}\label{eq:gap4b}
\frac{R(M)}{R^\ast(M)} \le 2(D+1)L\left[ 8(D+1)^2L^2 + 8\gamma(D+1)L \right] \le 32(D+1)^3L^3.
\end{equation}

\subsubsection{Gap}
By combining \eqref{eq:gap1}, \eqref{eq:gap2a}, \eqref{eq:gap2b}, \eqref{eq:gap3a}, \eqref{eq:gap3b}, \eqref{eq:gap4a} and \eqref{eq:gap4b}, we get:
\begin{equation*}
\frac{R(M)}{R^\ast(M)} \le 37(D+1)^3L^3.
\end{equation*}

\subsection{Proof of preliminary results}\label{app:gap-proof-prelims-proof}

\begin{proof}[Proof of Lemma~\ref{lemma:achievability}]
For $h\in H$, $\alpha_hM=0$, and thus, by Lemma~\ref{lemma:single-level}, $R_h(M)=\tilde R_h(0)=KU_h$.

For $j\in J$, $\alpha_jM=\frac{N_j}{d_j}$, and thus, by Lemma~\ref{lemma:single-level}, $R_j(M)=\tilde R_j(N_j/d_j)=0$.

For $i\in I'$, $\alpha_iM=\frac{\sqrt{N_iU_i}}{S_I}(M-T_J)$, and thus Lemma~\ref{lemma:single-level} gives:
\[
R_i(M) \le \frac{N_iU_i}{\alpha_iM}-d_iU_i = \frac{S_I\sqrt{N_iU_i}}{M-T_J} - d_iU_i.
\]
Furthermore, since $\alpha_iM\ge\frac{N_i}{K}$, we have:
\begin{IEEEeqnarray*}{rCl}
R_i(M) &=& \left(\frac{S_I\sqrt{N_iU_i}}{M-T_J}-d_iU_i\right)\left(1-\left(1-\frac{d_i\alpha_iM}{N_i}\right)^{K/d_i}\right)\\
&\ge& \frac{S_I\sqrt{N_iU_i}}{M-T_J}\left( 1 - \exp\left\{-\frac{K}{d_i}\cdot\frac{d_i\alpha_iM}{N_i}\right\} \right)\\
&\ge& \frac{S_I\sqrt{N_iU_i}}{M-T_J}(1-e^{-1}).
\end{IEEEeqnarray*}

Finally, for $i\in I''$, we have:
\[
\beta\frac{N_i}{d_i} \le \alpha_iM \le \frac{N_i}{d_i}.
\]
Moreover, we have $\tilde R_i(\beta N_i/d_i)\le\frac1\beta d_iU_i$, and $\tilde R_i(N_i/d_i)=0$.
Therefore, by the convexity of $\tilde R_i(\cdot)$, we get:
\[
R_i(M) \le \frac1\beta d_iU_i\left(1-\frac{\alpha_iM}{N_i/d_i}\right) = \frac1\beta d_iU_i\left(1-\frac{M-T_J}{\frac{1}{d_i}\sqrt{\frac{N_i}{U_i}}S_I}\right).
\]
Furthermore,
\[
R_i(M) \ge \left(\frac{N_iU_i}{\alpha_iM}-d_iU_i\right)(1-e^{-1}) = (1-e^{-1})\frac{N_iU_i}{\alpha_iM}\left(1-\frac{d_i\alpha_iM}{N_i}\right)
\ge (1-e^{-1})d_iU_i\left(1-\frac{M-T_J}{\frac{1}{d_i}\sqrt{\frac{N_i}{U_i}}S_I}\right).
\]
\end{proof}

\begin{proof}[Proof of Lemma~\ref{lemma:level-spacing}]
Assume, for the sake of contradiction, that there were two levels $i,j\in I''$. Without loss of generality, assume $i<j$, so that $U_i/N_i>U_j/N_j$. Then, by \eqref{eq:m-conditions-i1}, we have:
\[
\frac{\beta}{d_i}\sqrt{\frac{N_i}{U_i}}
\le \frac{\beta}{d_j}\sqrt{\frac{N_j}{U_j}}
< \tilde M
\le \sqrt{\frac{N_i}{U_i}}
\le \sqrt{\frac{N_j}{U_j}}.
\]
However, this implies, using \eqref{eq:beta}:
\[
\frac{\sqrt{U_i/N_i}}{\sqrt{U_j/N_j}} < \frac{d_j}{\beta} \le D\cdot16\gamma(D+1)L \le 16\gamma(D+1)^2L = q_0,
\]
thus contradicting the regularity condition.
\end{proof}

\begin{proof}[Proof of Lemma~\ref{lemma:uba}]
Let $i\in H\cup I'$.

If $i\in H$, then:
\[
\frac{\beta}{d_{i_1}}\sqrt{\frac{N_{i_1}}{U_{i_1}}} \le \frac{M-T_J}{S_I} \le \frac1K\sqrt{\frac{N_i}{U_i}}
\implies R_i(M) \le KU_i \le \frac{d_{i_1}}{\beta}\sqrt{\frac{N_iU_iU_{i_1}}{N_{i_1}}}.
\]

If $i\in I'$, then:
\[
\frac{M-T_J}{S_I} \ge \frac{\beta}{d_{i_1}}\sqrt{\frac{N_{i_1}}{U_{i_1}}}
\implies R_i(M) \le \frac{S_I\sqrt{N_iU_i}}{M-T_J} \le \frac{d_{i_1}}{\beta}\sqrt{\frac{N_iU_iU_{i_1}}{N_{i_1}}}.
\]
\end{proof}

\begin{proof}[Proof of Lemma~\ref{lemma:lbb}]
Use Lemma~\ref{lemma:converse-general}, with $l=i^\ast$, $A=J\cup\{i_1\}$, $s=\floor{\delta\frac{\sqrt{N_{i^\ast}/U_{i^\ast}}}{\sqrt{N_{i_1}/U_{i_1}}}}$, and $b=\floor{\frac{1}{\delta^2}\frac{N_{i_1}}{U_{i_1}}}$, where $\delta=\frac{1}{4L(D+1)}$.

First, note that $s\ge1$ since $q_0\ge\frac1\delta$, as well as $b\ge1$, and:
\begin{align*}
b &\ge \frac{1}{2\delta^2}\frac{N_{i_1}}{U_{i_1}} \ge \frac{N_{i_1}}{U_{i_1}} \ge \frac{N_j}{U_j},
\end{align*}
for all $j\in J$, and therefore $b\ge\frac{N_{i_1}}{d_{i_1}U_{i_1}}$ and $b\ge\frac{N_j}{d_jU_j}$.
This evaluates \eqref{eq:minimize-A}.
Furthermore:
\[
s^2b \le \frac{N_{i^\ast}}{U_{i^\ast}},
\]
thus evaluating \eqref{eq:minimize-l}.

Using these in the lower bound:
\begin{IEEEeqnarray*}{rCl}
R^\ast(M)
&\ge& \frac{1}{D+1}\left(\delta\sqrt{\frac{N_{i^\ast}U_{i_1}}{U_{i^\ast}N_{i_1}}} - d_i\right)U_{i^\ast} - \frac{M-T_J-\frac{N_{i_1}}{U_{i_1}}}{\frac{1}{2\delta^2}\frac{N_{i_1}}{U_{i_1}}}\\
&\ge& \sqrt{\frac{N_{i^\ast}U_{i^\ast}U_{i_1}}{N_{i_1}}}\left(\frac{\delta}{D+1} - \frac{1}{q_0} \right) - 2\delta^2 \cdot\frac{U_{i_1}}{N_{i_1}} \cdot \left( \frac{1}{d_{i_1}} \sqrt{\frac{N_{i_1}}{U_{i_1}}}S_I - \frac{N_{i_1}}{d_{i_1}} \right)\\
&\ge& \sqrt{\frac{N_{i^\ast}U_{i^\ast}U_{i_1}}{N_{i_1}}}\left(\frac{\delta}{D+1} - \frac{1}{q_0} - 2\delta^2L \right)\\
&\ge& \frac{1}{16L(D+1)^2}\sqrt{\frac{N_{i^\ast}U_{i^\ast}U_{i_1}}{N_{i_1}}}
\end{IEEEeqnarray*}
\end{proof}

\begin{proof}[Proof of Lemma~\ref{lemma:dominance}]
Assume first that $m\in H$.
Let $i\in I'$ be arbitrary.
By the definition of $m$, we have, from Lemma~\ref{lemma:achievability}:
\begin{IEEEeqnarray*}{r'C'rCl}
R_i(M) \le R_m(M)
&\implies& (1-e^{-1})\frac{S_I\sqrt{N_iU_i}}{M-T_J} &\le& KU_m\\
&\implies& \sqrt{N_iU_i} &\le& \frac{1}{1-e^{-1}}\cdot KU_m \cdot \frac{M-T_J}{S_I}\\
&\implies& \sqrt{N_iU_i} &\le& \frac{1}{1-e^{-1}}\cdot KU_m \cdot \frac{1}{K}\sqrt{\frac{N_m}{U_m}}\\
&\implies& \sqrt{N_iU_i} &\le& \frac{1}{1-e^{-1}}\cdot\sqrt{N_mU_m}.
\end{IEEEeqnarray*}
Therefore, $S_{I'}\le\gamma L\sqrt{N_mU_m}$.

Now assume that $m\in I'$.
Let $i\in I'$ be arbitrary.
By the definition of $m$, we have, from Lemma~\ref{lemma:achievability}:
\begin{IEEEeqnarray*}{r'C'rCl}
R_i(M) \le R_m(M)
&\implies& (1-e^{-1})\frac{S_I\sqrt{N_iU_i}}{M-T_J} &\le& \frac{S_I\sqrt{N_mU_m}}{M-T_J}\\
&\implies& \sqrt{N_iU_i} &\le& \frac{1}{1-e^{-1}}\cdot\sqrt{N_mU_m}.
\end{IEEEeqnarray*}
Therefore, $S_{I'}\le\gamma L\sqrt{N_mU_m}$.
\end{proof}

\subsection{Special case: small number of caches}\label{app:gap-proof-smallK}

Previously, we were looking at the case where $K\ge k_0$.
We will now study the opposite case: $K<k_0$.

When the number of caches is small, we use a different caching scheme to simplify the analysis.
First, we find the unique level $i^\ast$ such that:
\[
\sum_{i=1}^{i^\ast-1} \frac{N_i}{d_i}
\le M <
\sum_{i=1}^{i^\ast} \frac{N_i}{d_i}.
\]
Recall that the level numbers are ordered in decreasing popularity, \emph{i.e.}, $1$ is the most popular level and $L$ is the least popular.
Once $i^\ast$ is found, define the sets:
\begin{align*}
H &= \left\{ i^\ast+1,\ldots,L \right\};\\
I &= \left\{ i^\ast \right\};\\
J &= \left\{ 1,\ldots,i^\ast-1 \right\}.
\end{align*}
As before, we will fully store the set $J$ and give no memory to the set $H$.
The set $I$, however, will use the remaining memory and apply a conventional caching scheme, \emph{i.e.}, parts of files are stored and the user requests are served by supplying the remaining parts through multiple unicasts.
The resulting broadcast rate is:
\begin{equation}
\label{eq:smallK-ach}
R(M)
= \sum_{h\in H} KU_h
+ KU_{i^\ast}\left( 1 - \frac{M-T_J}{N_{i^\ast}/d_{i^\ast}} \right)
\le k_0\left[ LU_m
+ U_{i^\ast}\left( 1 - \frac{M-T_J}{N_{i^\ast}/d_{i^\ast}} \right)
\right],
\end{equation}
where
\(
m = \argmax_{h\in H} U_h
\)
is the level with the largest individual rate among the levels in $H$.

We will use Lemma~\ref{lemma:converse-general}, with $A=I\cup J$, $l=m$, $s=d_m$ and $b=\ceil{N_{i^\ast}/U_{i^\ast}}$.
Note that:
\begin{align*}
b &\ge \frac{N_{i^\ast}}{U_{i^\ast}} \ge \frac{N_j}{U_j},
\quad j\in J;\\
b &\le 2\frac{N_{i^\ast}}{U_{i^\ast}} \le 2\frac{N_h}{U_h},
\quad h\in H.
\end{align*}
Therefore:
\begin{IEEEeqnarray*}{rCl}
R^\ast(M)
&\ge& \frac{1}{D+1} \min\left\{ U_m , \frac{N_m}{b} \right\}
+ \sum_{i=1}^{i^\ast} \min\left\{ U_i , \frac{N_i}{b} \right\}\\
&\ge& \frac{1}{2(D+1)} U_m
+ \frac{N_{i^\ast}/d_{i^\ast} - (M-T_J)}{2N_{i^\ast}/U_{i^\ast}}\\
&=& \frac{1}{2(D+1)} U_m
+ \frac{1}{2d_{i^\ast}} U_{i^\ast} \left( 1 - \frac{M-T_J}{N_{i^\ast}/d_{i^\ast}} \right).
\IEEEyesnumber \label{eq:smallK-conv}
\end{IEEEeqnarray*}

Combining \eqref{eq:smallK-ach} with \eqref{eq:smallK-conv}, we get:
\[
\frac{R(M)}{R^\ast(M)}
\le 2Lk_0(D+1)
= 32(D+1)^3L(\gamma L+1).
\]

\section{Proof of the lower bounds on the optimal rate (Lemmas~\ref{lemma:converse-cutset} and~\ref{lemma:converse-general})}\label{app:converse-proof}

To prove the information-theoretic lower bounds, the following notation will be useful.
If there are $K$ APs numbered $1$ through $K$, then, for all $k\in\{1,\ldots,K\}$, we define $Z_k$ as the contents of the cache of the $k$-th AP.
Furthermore, we label the broadcast message as $X^\mathbf{r}$, where $\mathbf{r}=(r_1,\ldots,r_{KU})$ is a \emph{request vector} indicating what file is demanded by each of the $KU$ users.
Finally, for each level $i$, we define $\mathcal{W}^i$ as the set of files in level $i$. The individual files are denoted as $W^i_1,\ldots,W^i_{N_i}$.

\subsection{Proof of the cut-set bounds (Lemma~\ref{lemma:converse-cutset})}

Select $v$ users from level $i$, such that the total number of caches that are connected to at least one of these users is the smallest.
This results in $s=\ceil{v/U_i}+(d_i-1)$ caches.
We now send $b=\floor{N_i/v}$ different broadcasts $X^{\mathbf{r}_1},\ldots,X^{\mathbf{r}_b}$ aimed at $b$ different request vectors:
\begin{IEEEeqnarray*}{rCl}
\mathbf{r}_1 &=& (W^i_1,\ldots,W^i_v)\\
\mathbf{r}_2 &=& (W^i_{v+1},\ldots,W^i_{2v})\\
&\vdots& \\
\mathbf{r}_b &=& \left( W^i_{\left(\floor{N_i/v}-1\right)v+1},\ldots,W^i_{\floor{N_i/v}v} \right).
\end{IEEEeqnarray*}
These broadcast messages should thus allow the $v$ users to decode, together, $v\floor{N_i/v}$ files in total.
Therefore, we can write:
\begin{IEEEeqnarray*}{rCl}
bRF + sMF
&\ge& H\left( Z_1,\ldots,Z_s,X^{\mathbf{r}_1},\ldots,X^{\mathbf{r}_b} \right)\\
&=& H\left( Z_1,\ldots,Z_s,X^{\mathbf{r}_1},\ldots,X^{\mathbf{r}_b} \middle\vert W^i_1,\ldots,W^i_{v\floor{N_i/v}} \right)\\
&& {} + H\left(W^i_1,\ldots,W^i_{v\floor{N_i/v}}\right) - H\left(W^i_1,\ldots,W^i_{v\floor{N_i/v}} \middle\vert Z_1,\ldots,Z_s,X^{\mathbf{r}_1},\ldots,X^{\mathbf{r}_b} \right)\\
&\overset{(\ast)}{\ge}& H\left(W^i_1,\ldots,W^i_{v\floor{N_i/v}}\right) \cdot \left(1-\epsilon_F\right)\\
&=& v\floor{N_i/v} \cdot F \cdot \left(1-\epsilon_F\right),
\end{IEEEeqnarray*}
where $(\ast)$ follows from Fano's inequality.

As $F$ grows, this results in:
\begin{IEEEeqnarray*}{rCl}
bR + sM &\ge& v\floor{N_i/v}\\
\floor{N_i/v}R + \left(\ceil{v/U_i} + (d_i-1)\right)M &\ge& v\floor{N_i/v}\\
R &\ge& v - \frac{\ceil{v/U_i} + (d_i-1)}{\floor{N_i/v}}M.
\end{IEEEeqnarray*}
This concludes the proof.\qed

\subsection{Sliding window subset entropy inequality}
\newcommand{\defeq}{\overset{\text{def}}{=}}

To prove the non-cut-set bounds, we use an entropy inequality given in \cite{Liu14} called the sliding window subset entropy inequality.
We briefly present it in this section.

Define the following operator for any integers $m$ and $n$:
\begin{equation}
\langle m \rangle_n \defeq \left\{
\begin{aligned}
& m \bmod n \qquad & \text{if $m\bmod n\not=0$}\\
& n & \text{if $m\bmod n=0$}
\end{aligned}\right..
\end{equation}
The inequality is given in the following lemma.
\begin{lemma}[Sliding window subset entropy inequality {\cite[Theorem~3]{Liu14}}]
\label{lemma:sliding-window}
Let $(Y_1,\ldots,Y_n)$ be $n$ jointly distributed random variables.
Define, for simplicity, $\langle m\rangle\defeq\langle m\rangle_n$ for all integers $m$.
Then, for any $k\in\{1,\ldots,n-1\}$, we have:
\[
\frac{1}{k}\sum_{i=1}^n H\left(Y_i,\ldots,Y_{\langle i+k-1\rangle}\right)
\ge
\frac{1}{k+1}\sum_{i=1}^n H\left(Y_i,\ldots,Y_{\langle i+k\rangle}\right).
\]
\end{lemma}

\subsection{Proof of the non-cut-set bounds (Lemma~\ref{lemma:converse-general})}

For this proof, we make the following assumptions.
First, the level $l$ in the statement of the lemma is assumed to be of lower popularity than all the levels in the set $A$.
Indeed, in the gap analysis of Appendix~\ref{app:gap}, we always choose $l$ and $A$ this way.
Second, we assume that $d_1\le d_2\le \cdots\le d_L$.
As seen in Section~\ref{sec:numerics}, this is not an unreasonable assumption in most scenarios.
Therefore, $d_l\ge d_j$ for all $j\in A$.
We note that the lemma can be generalized, and a very similar result can be derived when these assumptions do not hold.

Consider $K$ sets of $b$ broadcast messages each, and denote the $i$-th such set by $\mathcal{X}^{(i)}$.
Then,
\begin{IEEEeqnarray*}{rCl}
sbR+sM &\ge& \frac{s}{K} \sum_{i=1}^K H\left( Z_i,\mathcal{X}^{(i)} \right).
\end{IEEEeqnarray*}

We choose the broadcast messages as follows.
Recall that, for every level $j\in A$, each user has access to the caches $(Z_i,\ldots,Z_{\langle i+d_j-1\rangle})$, for some $i\in\{1,\ldots,K\}$.
We say that these users are \emph{cache-indexed by $i$}.
The $d_j$ sets of broadcasts $\left(\mathcal{X}^{(i)},\ldots,\mathcal{X}^{(\langle i+d_j-1\rangle)}\right)$ are chosen so that the $U_j$ users cache-indexed by $i$ decode the first $w_j$ files of level $j$, where:
\[
w_j = \min\left\{ d_jbU_j , N_j \right\}.
\]
Correspondingly, we denote this set of files by: $\mathcal{\tilde W}^{(j)}$.
%Finally, we choose the broadcast such that all the level-$l$ users cache-indexed by $\{1,\ldots, s-d_l+1\}$ can collaboratively decode the first $w_l'$ files of level $l$ using the $sb$ broadcasts $(\mathcal{X}^{(1)},\ldots,\mathcal{X}^{(s)})$, where:
%\[
%w_l' = \min\left\{ s(s-d_l+1)bU_l , N_l \right\}.
%\]
The choice of the broadcasts with regards to the demands of the users in level $l$ will be decided afterwards.

Assume, without loss of generality, that $A=\{1,\ldots,t\}$, such that $d_1\le d_2\le\cdots\le d_t$.
We will use, in the following inequalities, Lemma~\ref{lemma:sliding-window} extensively, and will denote each step that uses it by $(\ast)$.
Steps that use Fano's inequality will be denoted by $(\ast\ast)$.
\begin{IEEEeqnarray*}{rCl}
sbR+sM &\ge& \frac{s}{K} \sum_{i=1}^K H\left( Z_i,\mathcal{X}^{(i)} \right)\\
&\overset{(\ast)}{\ge}& \frac{s}{K}\cdot\frac{1}{d_1} \sum_{i=1}^K H\left(Z_i,\mathcal{X}^{(i)},\ldots,Z_{i+d_1-1},\mathcal{X}^{(i+d_1-1)}\right)\\
&\overset{(\ast\ast)}{\ge}& \frac{s}{K}\cdot\frac{1}{d_1} \sum_{i=1}^K \left[ H\left(Z_i,\mathcal{X}^{(i)},\ldots,Z_{i+d_1-1},\mathcal{X}^{(i+d_1-1)} \middle\vert\mathcal{\tilde W}^1\right) + N_1(1-\epsilon_F) \right]\\
&=& \frac{s}{K}\cdot\frac{1}{d_1} \sum_{i=1}^K H\left(Z_i,\mathcal{X}^{(i)},\ldots,Z_{i+d_1-1},\mathcal{X}^{(i+d_1-1)} \middle\vert\mathcal{\tilde W}^1\right) + s\frac{N_1}{d_1}(1-\epsilon_F)\\
&\overset{(\ast)}{\ge}& \frac{s}{K}\cdot\frac{1}{d_2} \sum_{i=1}^K H\left(Z_i,\mathcal{X}^{(i)},\ldots,Z_{i+d_2-1},\mathcal{X}^{(i+d_2-1)} \middle\vert\mathcal{\tilde W}^1\right) + s\frac{N_1}{d_1}(1-\epsilon_F)\\
&\overset{(\ast\ast)}{\ge}& \frac{s}{K}\cdot\frac{1}{d_2} \sum_{i=1}^K \left[ H\left(Z_i,\mathcal{X}^{(i)},\ldots,Z_{i+d_2-1},\mathcal{X}^{(i+d_2-1)} \middle\vert\mathcal{\tilde W}^1,\mathcal{\tilde W}^2\right) + N_2(1-\epsilon_F) \right] + s\frac{N_1}{d_1}(1-\epsilon_F)\\
&=& \frac{s}{K}\cdot\frac{1}{d_2} \sum_{i=1}^K H\left(Z_i,\mathcal{X}^{(i)},\ldots,Z_{i+d_2-1},\mathcal{X}^{(i+d_2-1)} \middle\vert\mathcal{\tilde W}^1,\mathcal{\tilde W}^2\right) + s\left(\frac{N_1}{d_1}+\frac{N_2}{d_2}\right)(1-\epsilon_F)\\
&\ge& \cdots\\
&\ge& \frac{s}{K}\cdot\frac{1}{d_t}\sum_{i=1}^K H\left(Z_i,\mathcal{X}^{(i)},\ldots,Z_{i+d_t-1},\mathcal{X}^{(i+d_t-1)}\middle\vert\mathcal{\tilde W}^1,\ldots,\mathcal{\tilde W}^t\right) + s\left(\frac{N_1}{d_1}+\cdots+\frac{N_t}{d_t}\right)(1-\epsilon_F). \\*\IEEEyesnumber\label{eq:converse-step}
\end{IEEEeqnarray*}

Our next step is to combine the remaining entropy terms in the sum in a way that forms a cut-set bound for level $l$.
For that, we have to choose the level-$l$ demands that the broadcasts should serve.
To avoid any inconsistencies in the choice of the broadcasts, we have to consider two cases.

Let $\delta=\frac{1}{1+\frac{1}{D}}$, where $D$ is the largest degree.
First, if $s\ge\delta K$, then we choose the broadcasts such that the $(s-d_l+1)U_j$ users cache-indexed by the first $(s-d_l+1)$ caches can collaboratively decode the first $\min\left\{sb(s-d_l+1)U_j\right\}$ using the first $s$ sets of broadcasts.
Carrying out the calculations:
\begin{IEEEeqnarray*}{rCl}
&& \frac{s}{K}\cdot\frac{1}{d_t}\sum_{i=1}^K H\left(Z_i,\mathcal{X}^{(i)},\ldots,Z_{i+d_t-1},\mathcal{X}^{(i+d_t-1)}\middle\vert\mathcal{\tilde W}^1,\ldots,\mathcal{\tilde W}^t\right)\\
&\ge& \frac{s}{K}\cdot\frac{1}{d_t} H\left(Z_1,\mathcal{X}^{(1)},\ldots,Z_s,\mathcal{X}^{(s)}\middle\vert\mathcal{\tilde W}^1,\ldots,\mathcal{\tilde W}^t\right)\\
&\ge& \frac{\delta}{D} H\left(Z_1,\mathcal{X}^{(1)},\ldots,Z_s,\mathcal{X}^{(s)}\middle\vert\mathcal{\tilde W}^1,\ldots,\mathcal{\tilde W}^t\right)\\
&\overset{(\ast\ast)}{\ge}& \frac{\delta}{D} \min\left\{ s(s-d_l+1)bU_l , N_l \right\} (1-\epsilon_F).\\
&\ge& \frac{1}{D+1} \min\left\{ s(s-d_l+1)bU_l , N_l \right\} (1-\epsilon_F). \IEEEyesnumber\label{eq:cut-set-1}
\end{IEEEeqnarray*}

Second, if $s<\delta K$, then,
\begin{IEEEeqnarray*}{rCl}
&& \frac{s}{K}\cdot\frac{1}{d_t}\sum_{i=1}^K H\left(Z_i,\mathcal{X}^{(i)},\ldots,Z_{i+d_t-1},\mathcal{X}^{(i+d_t-1)}\middle\vert\mathcal{\tilde W}^1,\ldots,\mathcal{\tilde W}^t\right)\\
&\overset{(\ast)}{\ge}& \frac{s}{K}\cdot\frac{1}{s} \sum_{i=1}^K H\left(Z_i,\mathcal{X}^{(i)},\ldots,Z_{i+s-1},\mathcal{X}^{(i+s-1)}\middle\vert\mathcal{\tilde W}^1,\ldots,\mathcal{\tilde W}^t\right)\\
&\ge& \frac{1}{K} \sum_{i=1}^{K-s} H\left(Z_i,\mathcal{X}^{(i)},\ldots,Z_{i+s-1},\mathcal{X}^{(i+s-1)}\middle\vert\mathcal{\tilde W}^1,\ldots,\mathcal{\tilde W}^t\right).
\end{IEEEeqnarray*}
Because we now have only $K-s$ terms in the sum, and $s$ cache variables in each entropy term, we can choose broadcast message sets $\mathcal{X}^{(i)}$ such that all $(s-d_l+1)U_l$ users connected to $s$ consecutive caches can together decode up to $(s-d_l+1)U_l\cdot sb$ files, without worrying about inconsistencies.
Hence:
\begin{IEEEeqnarray*}{rCl}
&& \frac{1}{K} \sum_{i=1}^{K-s} H\left(Z_i,\mathcal{X}^{(i)},\ldots,Z_{i+s-1},\mathcal{X}^{(i+s-1)}\middle\vert\mathcal{\tilde W}^1,\ldots,\mathcal{\tilde W}^t\right)\\
&\overset{(\ast\ast)}{\ge}& \frac{1}{K} \cdot (K-s)\min\left\{ s(s-d_l+1)U_lb , N_l \right\} (1-\epsilon_F)\\
&\ge& (1-\delta) \min\left\{ s(s-d_l+1)U_lb , N_l \right\} (1-\epsilon_F).\\
&\ge& \frac{1}{D+1} \min\left\{ s(s-d_l+1)U_lb , N_l \right\} (1-\epsilon_F). \IEEEyesnumber\label{eq:cut-set-2}
\end{IEEEeqnarray*}

By combining \eqref{eq:cut-set-1} and \eqref{eq:cut-set-2} with \eqref{eq:converse-step}, we get the result of the lemma.\qed

\section{Proof of the rate achieved by ML-PAMA (Theorem~\ref{thm:multi-level-achievability})}\label{app:achievability}

Theorem~\ref{thm:multi-level-achievability} is a direct consequence of the following result.
\begin{lemma}\label{lemma:m-feasible}
For every $M\ge0$, there exists at least one $M$-feasible partition $(H,I,J)$ of the set of levels.
\end{lemma}
Before we prove the lemma, we will show how it implies the result of the theorem.

Given that an $M$-feasible partition exists, then the PAMA choice of the $\alpha_i$ parameters results in the following individual rates:
\begin{IEEEeqnarray*}{lCl"rCl}
\forall h&\in& H,& R_h(M) &=& KU_h,\\
\forall i&\in& I,& R_i(M) &\le& \frac{S_I\sqrt{N_iU_i}}{M-T_J}-d_iU_i,\\
\forall j&\in& J,& R_j(M) &=& 0.
\end{IEEEeqnarray*}
See Lemma~\ref{lemma:achievability} for more details.

By combining this with \eqref{eq:total-rate}, we get:
\begin{IEEEeqnarray*}{rCl}
R(M)
&=& \sum_{i=1}^L R_i(M)\\
&=& \sum_{h\in H} R_h(M) + \sum_{i\in I} R_i(M) + \sum_{j\in J} R_j(M)\\
&\le& \sum_{h\in H} KU_h + \sum_{i\in I} \left( \frac{S_I\sqrt{N_iU_i}}{M-T_J} - d_iU_i \right) + \sum_{j\in J} 0\\
&=& \sum_{h\in H} KU_h + \frac{S_I^2}{M-T_J}\\
&=& \sum_{h\in H} KU_h + \frac{\left(\sum_{i\in I}\sqrt{N_iU_i}\right)^2}{M-\sum_{j\in J}N_j/d_j},
\end{IEEEeqnarray*}
which is the rate expression stated in the theorem.

We will now prove Lemma~\ref{lemma:achievability}, thus completing the proof of Theorem~\ref{thm:multi-level-achievability}.

\begin{proof}[Proof of Lemma~\ref{lemma:achievability}]
We prove the existence of an $M$-feasible partition for each $M$ by construction.
We give an algorithm that constructs such a partition for all $M\ge0$ in $\Theta(L^2)$, shown in Algorithm~\ref{alg:m-feasible}.

We first observe that the inequalities associated with an $M$-feasible partition can be rewritten as:
\begin{IEEEeqnarray*}{rCl"rCcCl}
\forall h&\in&H, & && M &\le& f^{I,J}\left(\tilde m_h\right) + \frac{N_h}{K};\\
\forall i&\in&I, & f^{I,J}\left(\tilde m_i\right) &\le& M &\le& f^{I,J}\left(\tilde M_i\right);\\
\forall j&\in&J, & f^{I,J}\left(\tilde M_j\right) &\le& M,
\end{IEEEeqnarray*}
where, for any level $i$, we define:
\begin{IEEEeqnarray*}{rCl}
\tilde m_i &=& \frac{1}{K} \sqrt{\frac{N_i}{U_i}};\\
\tilde M_i &=& \frac{1}{d_i} \sqrt{\frac{N_i}{U_i}},
\end{IEEEeqnarray*}
and, for any subsets $A,B\subseteq\{1,\ldots,L\}$, we define:
\[
f^{A,B}(x) = x\cdot S_A + T_B.
\]

\begin{algorithm}
\caption{Algorithm that finds an $M$-feasible partition for all $M$.}
\label{alg:m-feasible}
\begin{algorithmic}[1]

\Procedure{PartitionLevels}{$K,\{N_i,U_i,d_i\}_{i=1}^L$}
\State Sort the terms $\left\{\tilde m_i, \tilde M_i\right\}_{i=1}^L$ and label the result as $(x_1,\ldots,x_{2L})$.

\State $I_0\gets\emptyset$
\State $J_0\gets\emptyset$
\For{$t\in\{1,\ldots,2L\}$}
  \If{$x_t=\tilde m_i$ for some $i$}
    \State $I_t \gets I_{t-1} \cup \{i\}$
    \State $J_t \gets J_{t-1}$
  \ElsIf{$x_t=\tilde M_i$ for some $i$}
    \State $I_t \gets I_{t-1} \backslash \{i\}$
    \State $J_t \gets J_{t-1} \cup \{i\}$
  \EndIf
  \State $Y_t \gets f^{I_{t-1},J_{t-1}}(x_t)$
\EndFor
\State $Y_{2L+1} \gets \infty$
\State Store $(Y_1,\ldots,Y_{2L+1})$ and $(I_1,J_1,\ldots,I_{2L},J_{2L})$.
\EndProcedure

\Procedure{GetPartition}{$M$}
\State Find $t$ such that $Y_t\le M < Y_{t+1}$
\State $(I,J) \gets (I_t,J_t)$
\State $H \gets \left(I_t\cup J_t\right)^c$
\State \Return $(H,I,J)$
\EndProcedure

\end{algorithmic}
\end{algorithm}

The first thing to note is that $f^{I,J}(\cdot)$ is an increasing function for any fixed $I$ and $J$. Therefore,
\begin{equation}\label{eq:f-order-1}
f^{I,J}\left(\tilde m_i\right) \le f^{I,J}\left(\tilde M_i\right),
\end{equation}
for all levels $i$.

Second, for any level $i$, and any disjoint sets $I$ and $J$ such that $i$ is not an element of either $I$ or $J$, we have:
\begin{IEEEeqnarray*}{rCl}
f^{I,J}\left(\tilde m_i\right)
&=& \frac{1}{K}\sqrt{\frac{N_i}{U_i}} \cdot S_I + T_J\\
&\le& \frac{1}{K}\sqrt{\frac{N_i}{U_i}} \cdot \left(S_I+\sqrt{N_iU_i}\right) + T_J\\
&=& \frac{1}{K}\sqrt{\frac{N_i}{U_i}} \cdot S_{I\cup\{i\}} + T_J\\
&=& f^{I\cup\{i\},J}\left(\tilde m_i\right).\IEEEyesnumber\label{eq:f-order-2}
\end{IEEEeqnarray*}
Moreover,
\begin{IEEEeqnarray*}{rCl}
f^{I\cup\{i\},J}\left(\tilde M_i\right)
&=& \frac{1}{d_i}\sqrt{\frac{N_i}{U_i}} \cdot S_{I\cup\{i\}} + T_J\\
&=& \frac{1}{d_i}\sqrt{\frac{N_i}{U_i}} \cdot \left(S_I + \sqrt{N_iU_i} \right) + T_J\\
&=& \frac{1}{d_i}\sqrt{\frac{N_i}{U_i}} \cdot S_I + \frac{N_i}{d_i} + T_J\\
&=& \frac{1}{d_i}\sqrt{\frac{N_i}{U_i}} \cdot S_I + T_{J\cup\{i\}}\\
&=& f^{I,J\cup\{i\}}\left(\tilde M_i\right).\IEEEyesnumber\label{eq:f-order-3}
\end{IEEEeqnarray*}

By combining \eqref{eq:f-order-1}, \eqref{eq:f-order-2}, and \eqref{eq:f-order-3}, we can conclude that the function $f$ preserves the order of the $(x_1,\ldots,x_{2L})$ terms.
In particular, the algorithm always results in:
\[
Y_1 \le Y_2 \le \cdots \le Y_{2L}.
\]

The last thing to check is that the partition returned by the algorithm for any $M$ is indeed $M$-feasible.
Recall that, for a particular $M$, the algorithm finds $t$ such that $Y_t \le M < Y_{t+1}$, and then selects the partition $(H,I,J)$ based on $I_t$ and $J_t$.
Because of the perfect matching between the $x_t$'s and the $Y_t$'s enabled by the function $f$, we can deduce some conditions on $M$.
First, all levels $i\in I$ would only be in the set $I$ because the algorithm passed by a $\tilde m_i$ term prior to reached $x_t$, but did not pass by the term $\tilde M_i$ yet.
Hence, $\tilde m_i \le x_t < x_{t+1} \le \tilde M_i$.
The perfect matching between $x$ and $Y$ gives:
\[
\frac{1}{K}\sqrt{\frac{N_i}{U_i}}S_I + T_J = f^{I_t,J_t}\left(\tilde m_i\right) \le Y_t \le M < Y_{t+1} \le f^{I_t,J_t}\left(\tilde M_i\right) = \frac{1}{d_i}\sqrt{\frac{N_i}{U_i}}S_I + T_J
\]
Similarly, for $h\in H$,
\[
M < Y_{t+1} \le f^{I_t,J_t}\left(\tilde m_h\right) \le \frac{1}{K}\sqrt{\frac{N_h}{U_h}}S_I + T_J + \frac{N_h}{K},
\]
and, for $j\in J$,
\[
\frac{1}{d_j}\sqrt{\frac{N_j}{U_j}}S_I+T_J = f^{I_t,J_t}\left(\tilde M_j\right) \le Y_t \le M,
\]
which are the conditions of $M$-feasibility, thus concluding the proof.
\end{proof}

%%fakesection
%\section{Proof of the single-level achievable rate (Lemma~\ref{lemma:single-level})}\label{app:single-level-proof}
%\myinput{ext-single-level-proof.tex}

%\section{Multi-level setup with a single user per cache}\label{app:single-user}

\end{document}